\documentclass[twocolumn]{aastex6}

\usepackage{xcolor}
\usepackage{amsmath}
\usepackage{affils}

\newcommand{\ghsc}[0]{$g_{\rm HSC}$}
\newcommand{\ihsc}[0]{$i_{\rm HSC}$}
\newcommand{\masq}[0]{mag arcsec$^{-2}$}

\newcommand{\gi}[0]{$\Delta (g-i)$}
\newcommand{\redshells}[0]{``red shell'' galaxies}
\newcommand{\smass}[0]{\log ( M_\star / M_\odot )}
\newcommand{\ssim}[0]{\! \sim \!}
\newcommand{\ftid}[0]{$f_{\rm tidal}$}
\newcommand{\ctf}[0]{$c_{\rm tf}$}
\newcommand{\mctf}[0]{c_{\rm tf}}
\newcommand{\nshells}[0]{214}
\newcommand{\nstreams}[0]{987}
\newcommand{\ntf}[0]{1,201}
\newcommand{\nsdss}[0]{21,208}
\newcommand{\ncolshells}[0]{78}
\newcommand{\ncolstreams}[0]{497}
\newcommand{\ncol}[0]{575}

\begin{document}
\shortauthors{Kado-Fong et al.}

\title{tidal features at $0.05<\MakeLowercase{z}<0.45$  in the Hyper Suprime-Cam Subaru Strategic Program: properties and formation channels}

\DeclareAffil{princeton} {Department of Astrophysical Sciences, Princeton University,
Princeton, NJ 08544, USA}
\DeclareAffil{columbia}{Department of Astronomy, Columbia University, 550 W 120th St, New York, NY 10027, USA}
\DeclareAffil{naoj}{National Astronomical Observatory of Japan, 2-21-1 Osawa, Mitaka, Tokyo 181-8588, Japan}
\DeclareAffil{subaru}{Subaru Telescope, NAOJ, 650 North A'ohoku Place, Hilo, HI 96720, USA}
\DeclareAffil{ipmu}{Kavli-IPMU, The University of Tokyo Institutes for Advanced Study, the University of Tokyo, Kashiwa 277-8583, Japan}
\DeclareAffil{santacruz}{Department of Astronomy and Astrophysics, University of California Santa Cruz, 1156 High St., Santa Cruz, CA 95064, USA}
\DeclareAffil{sokendai}{Graduate University for Advanced Studies (SOKENDAI), 2-21-1 Osawa, Mitaka, Tokyo 181-8588, Japan}

\affilauthorlist{
  E. Kado-Fong\affils{princeton},
  J. E. Greene \affils{princeton},
  D. Hendel \affils{columbia},
  A. M. Price-Whelan \affils{princeton},
  J. P. Greco \affils{princeton},
  A. D. Goulding \affils{princeton},
  S. Huang \affils{ipmu,santacruz},
  K. V. Johnston \affils{columbia},
  Y. Komiyama \affils{naoj,sokendai}
  C.-H. Lee \affils{subaru},
  N. B. Lust \affils{princeton},
  M. A. Strauss \affils{princeton},
  M. Tanaka \affils{naoj}
  }
  
  \date{\today}

\begin{abstract}

We present \ntf{} galaxies at $0.05<z<0.45$ that host tidal features, detected from the first $\ssim 200$ deg$^2$ of imaging from the Hyper Suprime-Cam Subaru Strategic Program (HSC-SSP). 
All galaxies in the present sample have spectroscopic observations from the Sloan Digital Sky Survey (SDSS) spectroscopic campaigns, generating a sample of \nsdss{} galaxies. Of these galaxies, we identify \nshells{} shell systems and \nstreams{} stream systems. For \ncol{} of these systems, we are additionally able to measure the $(g-i)$ colors of the tidal features. We find evidence for star formation in a subset of the streams, with the exception of streams around massive ellipticals, and find that stream host galaxies span the full range of stellar masses in our sample. Galaxies which host shells are predominantly red and massive: we find that observable shells form more frequently around ellipticals than around disc galaxies of the same stellar mass. Although the majority of the shells in our sample are consistent with being formed by minor mergers, $15\% \pm 4.4\%$ of shell host galaxies have $(g-i)$ colors as red as their host galaxy, consistent with being formed by  major mergers. These \redshells{} are additionally preferentially aligned with the major axis of the host galaxy, as previously predicted from simulations. We suggest that although the bulk of the observable shell population originates from fairly minor mergers, which preferentially form shells that are not aligned with the major axis of the galaxy, major mergers produce a significant number of observable shells. 
\end{abstract}


\section{Introduction}
In the hierarchical merging model of galaxy formation, galaxy mergers are a crucial mechanism through which galaxies grow in size and mass \citep{toomre1977, white1978, white1991}. This picture of galaxy formation has received strong support from modern cosmological simulations and semi-analytical models. A general consensus has emerged that for the highest masses, the growth of galaxies is dominated by matter accreted from other systems, with the \textit{ex situ} (accreted) stellar mass fraction reaching $\ssim 0.80$ \citep{oser2010, lee2013, rodriguezgomez2016}, suggesting that mergers are necessary to produce galaxies of $\smass \gtrsim 11.0$ \citep{lee2017}. 

The idea that mergers can generate extended features is well-established, having been shown first by \cite{pfleiderer1963} and \cite{toomre1972} for on-going mergers between two equal-mass discs. It is now also widely accepted that the extended, low surface brightness features found around low redshift galaxies are the product of past merging events \citep{malin1983, quinn1983, dupraz1986, fort1986, mihos1998, bartoskova2011, pop2017a}, and that the structure and characteristics of these features hold a substantial amount of information regarding the dynamics and assembly history of their hosts \citep[e.g.][]{johnston2008, belokurov2017}. 

These tidal features can be separated into two broad classes: ``shells'', which are characterized by umbrella-like fans and caustics of stars whose radius of curvature points towards the host galaxy, and ``streams'', which are extended ribbon-like structures near the host galaxy. The boundary between these two classes is intrinsically soft, as stream-like features can, due to line-of-sight projection effects, appear shell-like \citep{foster2014, greco2018} and have been shown to evolve into shell-like systems \citep{hendel2015}. Nevertheless, the ensemble characteristics of these shells and streams are distinct -- shells tend to be found around massive ellipticals and do not appear to host significant star formation \citep{malin1983, tal2009, atkinson2013}, while streams can be quite blue, and often appear around disc galaxies \citep{adamo2012, knierman2012, atkinson2013, higdon2014}. 

The nature of the mergers that formed these tidal features, however, is still debated. In the minor merger picture of shell formation, a low mass satellite falls radially into the potential of a more massive host and the stars of the satellite form the resultant tidal feature \citep[see, e.g., ][]{johnston2001, kawata2006, sanderson2013, hendel2015}. This mechanism of tidal feature formation has been successfully used to model many nearby systems (e.g. the Umbrella galaxy by \citealt{foster2014}, the Ophiuchus Stream by \citealt{pricewhelan2016}, Sumo Puff by \citealt{greco2018}), and the inferred merger mass ratios for most known tidal features that have been modeled are in agreement with this picture \citep{kawata2006,feldman2008,gu2013,foster2014}.

However, a separate formation channel has been proposed wherein the shells are formed by major mergers. Following \cite{hernquist1992}, it was recently reported by \cite{pop2017a} that the shells found around the most massive galaxies in the \textit{Illustris} simulation \citep{vogelsberger2014} are generated by major mergers, with the distribution of shell-forming events peaking at mergers between galaxies of approximately equal stellar mass. Observations of three dwarf galaxies in the Virgo Cluster \citep{paudel2017}, and of nine nearby early-type galaxies (ETGs) taken from the \cite{malin1983} sample \citep{carlsten2017} show signs of shells that are thought to be generated by major mergers. Unlike tidal features formed from minor mergers between a quiescent host and quiescent satellite, where the tidal feature is bluer than the core of its host galaxy due to the color-mass relation along the red sequence, these major merger shells should show metallicities close to those at the core of the galaxy \citep{pop2017b}, implying colors close to that of the core of the host \citep[see, e.g.][]{gallazzi2006}. This provides an observationally accessible difference between the minor and major merger scenarios.

The picture for stream formation is somewhat clearer: they are generally accepted to originate from infalling satellites with a relatively high angular momentum, although there is a smooth transition from stream to shell in the minor merger picture \citep[see, e.g.][]{hernquist1988, hendel2015}. Blue, low surface brightness tidal streams can, however, be caused by both recent minor \citep{knierman2012, knierman2013} and older major mergers \citep{rodruck2016}. In addition, there are red, quiescent streams which show no active star formation. These could be either the result of a dry merger or from quenching of a gas-rich satellite upon infall \citep{feldman2008}.

\begin{figure*}
\center{\includegraphics[width=\linewidth]{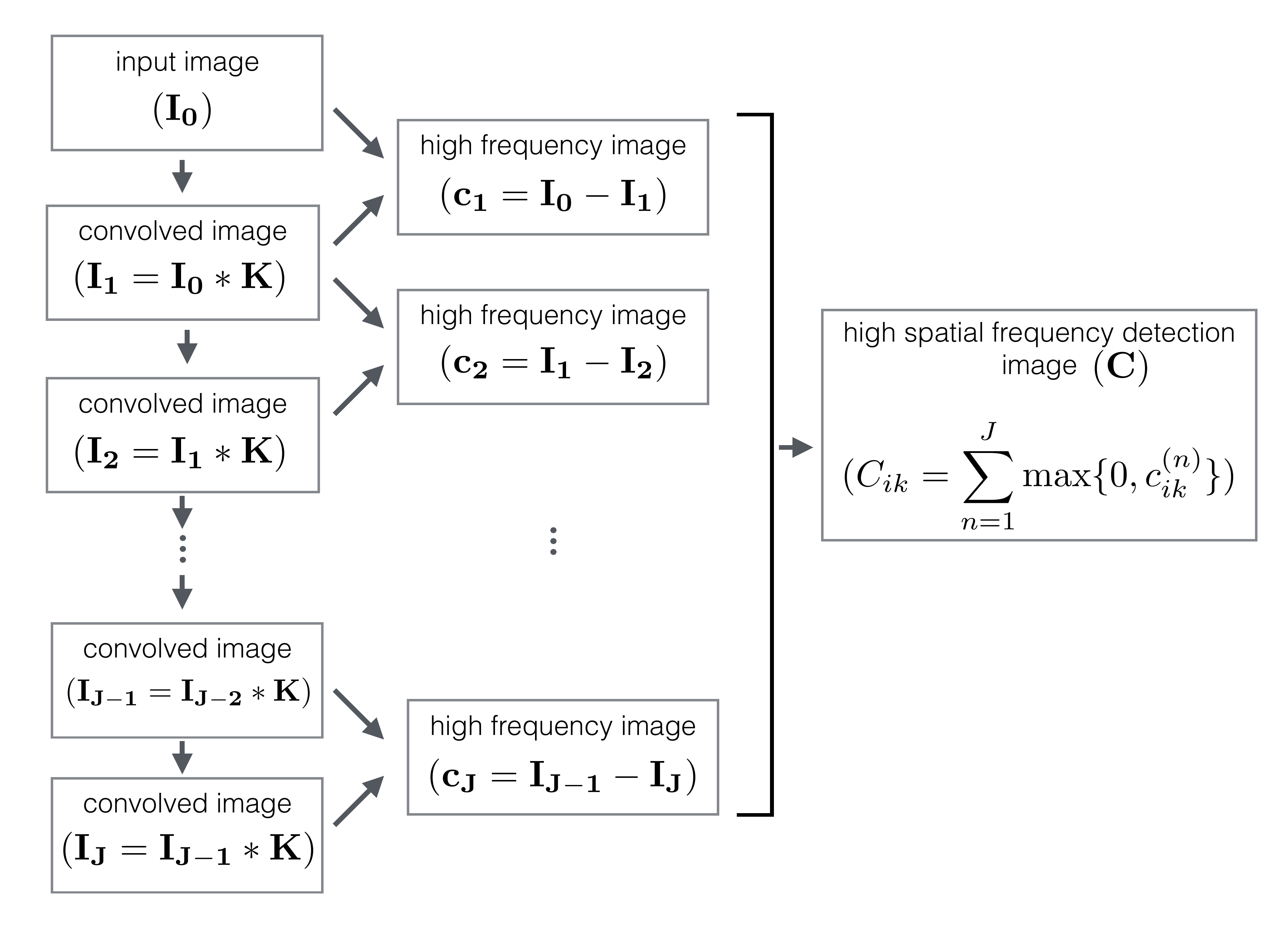}}
\caption{
A schematic diagram of our algorithm to detect tidal features in HSC-SSP data. From left to right, the flow chart shows the convolution step, generation of high spatial frequency images, and detection image creation.
}
\label{methodschematic}
\end{figure*}

The majority of observational work on tidal features in external galaxies to date has been based upon targeted observations of a relatively small number of galaxies \citep[see, e.g.][]{schweizer1988, martinezdelgado2009, martinezdelgado2010, kim2012, knierman2012, knierman2013, beaton2014, martinezdelgado2015, rodruck2016, greco2018}. Amongst those works that are able to derive tidal feature occurrence fractions, where \ftid{} is the fraction of all galaxies that host tidal features, observed values of \ftid{} vary by a factor of 7 \citep{malin1983, vandokkum2005, tal2009, nair2010, atkinson2013, hood2018}. This is due in part to variations in targeting strategy and detection methods, as well as from differences in imaging depth -- the majority of tidal features are expected to have a peak surface brightness of higher than $\ssim 30$ \masq{} \citep{johnston2008}. Moreover, the vast majority of the existing samples of tidal features are created via visual inspection.

The Hyper Suprime Cam Subaru Strategic Program (hereafter HSC-SSP) is an ongoing wide-field survey executed by the Hyper Suprime Cam mounted on the 8.2 meter Subaru telescope \citep[][Kawanomoto et al. 2018, in prep.]{miyazaki2012, furusawa2018, komiyama2018, miyazaki2018}. HSC-SSP is set to cover over $1400 ~{\rm deg}^2$ of the sky in five bands ($grizy$) to \ihsc{} $\ssim 26$ \citep[$5\sigma$ point source depth, see][]{aihara2018b,aihara2018}. The survey boasts a median $0.56''$ seeing in the \ihsc{} band \citep{aihara2018}, making it an optimal imaging campaign in which to detect extended, low surface brightness features around galaxies out to intermediate redshifts \citep[see][]{greco2017,greco2018, huang2018a,huang2018b}.

In this paper, we consider \nsdss{} galaxies with spectroscopic observations from the Sloan Digital Sky Survey DR12 \citep{alam2015} in the footprint of the Hyper Suprime-Cam Subaru Strategic Program (HSC-SSP) Wide layer  \citep{aihara2018}. The HSC Wide layer covers the largest on-sky area at a relatively shallow depth ($i\ssim 26$) relative to the Deep and UltraDeep Layers ($i\ssim27$ and $i\ssim28$, respectively). We describe the survey and sample selection in \autoref{sect_sampleselection}. Our aims are as follows. First, for a survey as large as HSC-SSP, it is necessary to construct an algorithm which is capable of producing a homogeneously selected sample of galaxies with tidal features. In \autoref{sect_algorithm}, we create an algorithm for automatic detection and, because there does not exist a suitable training set, we classify the morphologies of detected tidal feature through visual inspection. In \autoref{sect_results}, we compare the results of our method to previous, purely visual samples , and compare the nature of the stream and shell host galaxies to each other, as well as to the non-interacting galaxies in our sample. We present the results of measurements of the color of the tidal features in \autoref{sect_colormeas}. Finally, we discuss the demographics of the tidal feature hosts, and typical origins of the features in our sample. 

In this work, we adopt a cosmology of $\Omega_m=0.3$, $\Omega_\Lambda=0.7$, and $H_0 = 70$ km s$^{-1}$ Mpc$^{-1}$.

\section{Sample selection}\label{sect_sampleselection}
\subsection {HSC-SSP}
 In this work, we use the coadded images produced by the HSC software pipeline \textsf{hscPipe} \citep{bosch2018}. For extended, low surface brightness structure, accurate sky subtraction is crucial. \textsf{hscPipe} performs sky subtraction via fitting a sixth-order 2D Chebyshev polynomial to a binned average of the image (with a bin size of $128\times128$ pixels). Sky subtraction is performed prior to source detection, after initial source detection, and after final (per-visit) source detection during CCD processing. During coaddition, a constant background is measured and subtracted from the images. For a detailed description of the HSC-SSP data reduction process, see \cite{bosch2018}. Because the \ihsc{} band has the best seeing on average across HSC-SSP, we use the \ihsc{} band images as our detection images. 


We set our redshift limits at $0.05<z<0.45$, selecting from cross matches between the HSC Wide catalog from \textsf{hscPipe} (internal data release S16A, $\ssim 200$ deg$^2$) and the SDSS spectroscopic surveys. At the lowest redshift, we are currently unable to accurately detect tidal features due to issues in background subtraction around very extended galaxies (however, it is relevant to note that this issue has been corrected for future data releases). At high redshift, we lose the ability to detect tidal features due to cosmological surface brightness dimming and a loss of spatial resolution; a negligible number of galaxies at $z>0.45$ have automatically detected tidal features using our algorithm (see \autoref{occurrencefractions}). In \autoref{sect_detrecovery}, we investigate our surface brightness limits by injecting simulated tidal features at various spatial extents into non-interacting galaxies from our sample.


\begin{figure*}
\center{\includegraphics[width=\linewidth]{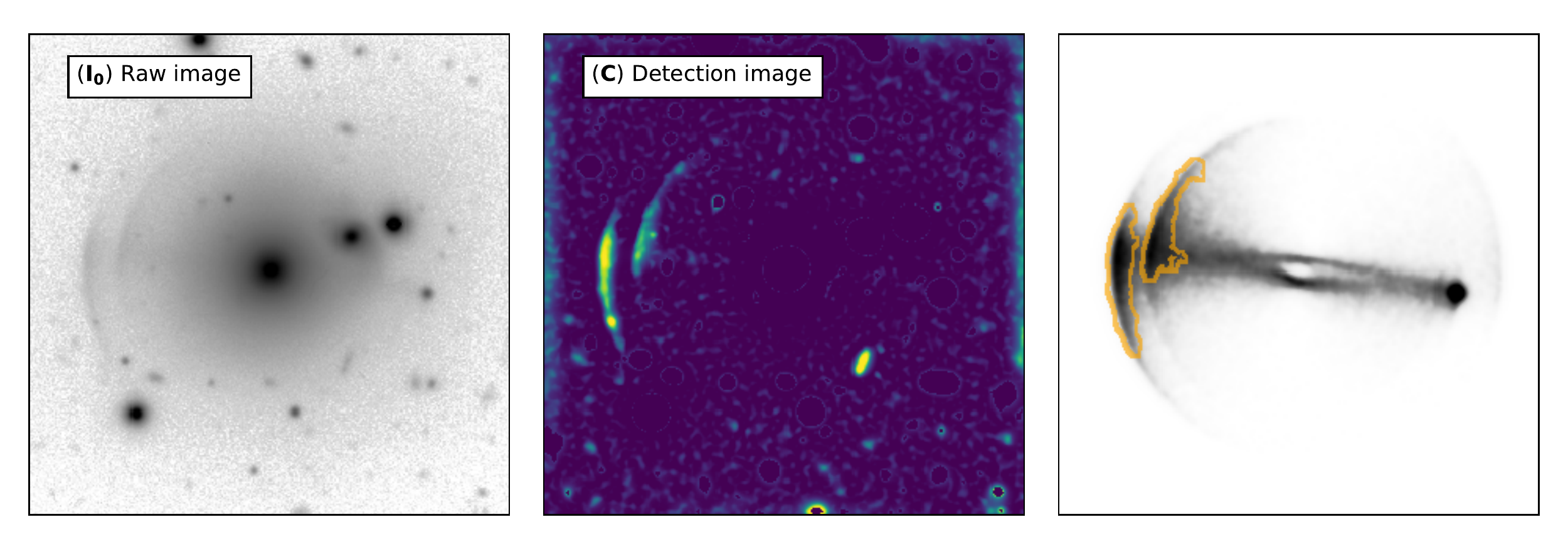}}
\caption{
An illustration of the filtering algorithm as applied to the sum of a non-interacting HSC galaxy and a simulated shell system from \cite{hendel2015} with a mean surface brightness of $\bar \mu_{i} = 25.4 $ \masq{}. \textit{Left:} $\mathbf{I_0}$, the sum of the non-interacting HSC galaxy cutout and simulated, PSF-convolved tidal feature system. \textit{Middle:} The detection map, $\mathbf{C}$, generated from the left panel. The two brightest caustics in the shell system are clearly visible (yellow-green regions, left of panel), along with a background galaxy. \textit{Right:} The simulated tidal feature from \cite{hendel2015}. The orange outlines show the boundaries of the region that our method identifies as a tidal feature candidate. All panels are shown with a logarithmic stretch.
}
\label{detex}
\end{figure*}

\subsection {SDSS spectra}

All of the galaxies in the present sample are selected to have spectra from SDSS DR12, and thus have known spectroscopic redshifts \citep{alam2015}. We additionally use stellar masses derived by \cite{chen2012} by fitting the SDSS spectra to principal components constructed from the models of \cite{bruzual2003}. We use the stellar masses calculated by \cite{kauffmann2003} when a stellar mass measurement from \cite{chen2012} is not available, where an offset calibrated from the galaxies where a stellar mass was available from both methods is applied to the stellar masses from \cite{kauffmann2003}. Both stellar mass estimates adopt a \cite{kroupa2001} initial mass function.

We include galaxies from both the SDSS Legacy catalog \citep[complete to $r<17.77$ mag,][]{strauss2002} and the Baryon Oscillation Spectroscopic Survey of SDSS-III \citep[BOSS, color-selected and roughly stellar mass complete, ][]{dawson2013, reid2016}. This generates a sample composed primarily of massive, low-$z$ galaxies with a bias towards elliptical galaxies. To avoid issues concerns about inhomogeneity of the SDSS samples, we consider our sample in bins of stellar mass and redshift. We additionally note that for $z<0.15$ (which we will consider as a low-$z$ subset of the full spectroscopic sample), the sample is dominated by galaxies from the SDSS Legacy catalog. We use the HSC catalog associated with the internal data release S16A to cross match HSC objects to their SDSS spectroscopic counterparts, resulting in \nsdss{} galaxies.

\section{Identification and characterization of tidal features }\label{sect_algorithm}

\subsection {Detection}\label{sect_detection}
Visually classifying every nearby galaxy in the HSC footprint is neither scalable nor feasible. It is therefore of interest to use this initial spectroscopic sample to develop an automated detection method that can be applied to the full HSC-SSP survey area.
However, an automated method for detecting tidal features should be able to operate in a single band (both to maximize the area over which the algorithm can function, and so that the algorithm can be applied to any arbitrary bandpass), agnostic to the global morphological characteristics of both the tidal features and host, and functional in the low surface brightness regime. To this end, we have developed a method that separates high spatial frequency features (e.g. streams, shell caustics) from low spatial frequency features (e.g. host galaxy light) in a single band image. 

Our algorithm is represented schematically in \autoref{methodschematic}. We iteratively separate low and high spatial frequency features, using the \ihsc{} band image output by \textsf{hscPipe}. We take our initial image as a $512 \times 512$ pixel cutout around a galaxy of interest in our parent sample ($\mathbf{I_0}$). The $n^{\rm th}$ image is then given by the convolution of the previous image and the kernel, $\mathbf{K}$ (leftmost column, \autoref{methodschematic}):

\begin{equation}
\mathbf{I_n} = \mathbf{I_{n-1}} \ast \mathbf{K},
\end{equation}

\noindent{}where $\mathbf{K}(x,y)=\mathbf{\phi}(x)^T\mathbf{\phi}(y)$. We choose $\mathbf{\phi}$ to be the one-dimensional B3-spline, using $\mathbf{\phi}(x)^T =  (\frac{1}{16} ~\frac{1}{4}~ \frac{3}{8}~ \frac{1}{4}~ \frac{1}{16} )$ pixels (where each pixel spans 0.168 arcsec), following \cite{starck2007}. We then isolate the high spatial frequency features in the image by constructing a high spatial frequency image, $\mathbf{c^{(n)}}$, from the difference of the smoothed and un-smoothed images. This is shown in the middle column of \autoref{methodschematic} as 
\begin{equation}
\mathbf{c^{(n)}} = \mathbf{I_{n-1}} -  \mathbf{I_n}.
\end{equation}
Thus, those features that have spatial frequencies which are present in $\mathbf{I_{n-1}}$ but removed in $\mathbf{I_{n}}$ have positive values in $\mathbf{c^{(n)}}$; $\mathbf{c^{(n)}}$ is essentially the output of the well-known unsharp masking algorithm on image $\mathbf{I_{n-1}}$ \citep[astronomical applications, see, e.g.][]{malin1977,meaburn1980}. 

We then construct a detection image of the high spatial frequency components in the image by stacking the positive components of each $\mathbf{c^{(n)}}$: 

\begin{equation} \label{detequation}
C_{ik} = \sum_{n=2}^{J} \max \{ 0, c^{(n)}_{ik} \},
\end{equation}

Where $C_{ik}$ is the pixel at ($i$,$k$) in the stacked image, $c^{(n)}_{ik}$ is the pixel at ($i$,$k$) in $\mathbf{c^{(n)}}$, and $J$ is the number of layers in the filter bank (rightmost column, \autoref{methodschematic}). We exclude the first component, $\mathbf{c}^{(1)}$, as it is dominated by noise features. 

For arbitrarily high $J$, the spatial frequencies probed by $c^{(J)}$ become arbitrarily low and can encompass the host galaxy. Similarly, the spatial frequencies associated with the host galaxy are strongly dependent on the physical size and redshift of the galaxy. The value of $J$ is therefore set somewhat heuristically. Here, we set $J=36$ (i.e. where the signal from one pixel can spread over at most a square with sides of 145 pixels), as the range of on-sky sizes for the galaxies in our sample, which tend to be massive and low redshift, allows us to adopt a static value for $J$.

\subsection{Contaminant removal}
This method to produce the detection image $\mathbf{C}$ generates high pixel values at the locations of tidal features, the cores of sources, spiral arms, and imaging artifacts. To remove contaminants from imaging artifacts, we reject detections where a saturated star is located within the image cutout, when greater than 20\% of the pixels in the cutout are flagged as missing data, or when greater than 95\% of the pixels in the cutout are flagged as being near a bright star. These cuts are tuned heuristically to remove imaging artifacts that produce high spatial frequency features.


We take the following approach to remove contamination from spiral arms and neighboring galaxies. First, a flux threshold is applied to identify the cores of the brightest objects. The core of the target galaxy is not flagged in this image. The threshold is then iteratively lowered: at each iteration, the detected regions (i.e., a group of contiguous pixels) that are either connected to a previously detected source or are considered circular ($| \pi R_{\rm obj}^2 - A_{\rm obj} | < 0.25$, where $R_{\rm obj}$ and $A_{\rm obj}$ are half the extent of the object and the area of the object in pixels, respectively) are flagged as non-target objects. For an application of this technique to low surface brightness galaxies see \cite{greco2017}. To remove detections from imaging artifacts and noise, we require that candidate detections cover at least $250$ pixels (approximately $7.05$ arcsec$^2$). For a visualization of our decomposition, cleaning, and detection process, see \autoref{appendix}.

\begin{figure}
\center{\includegraphics[width=\linewidth]{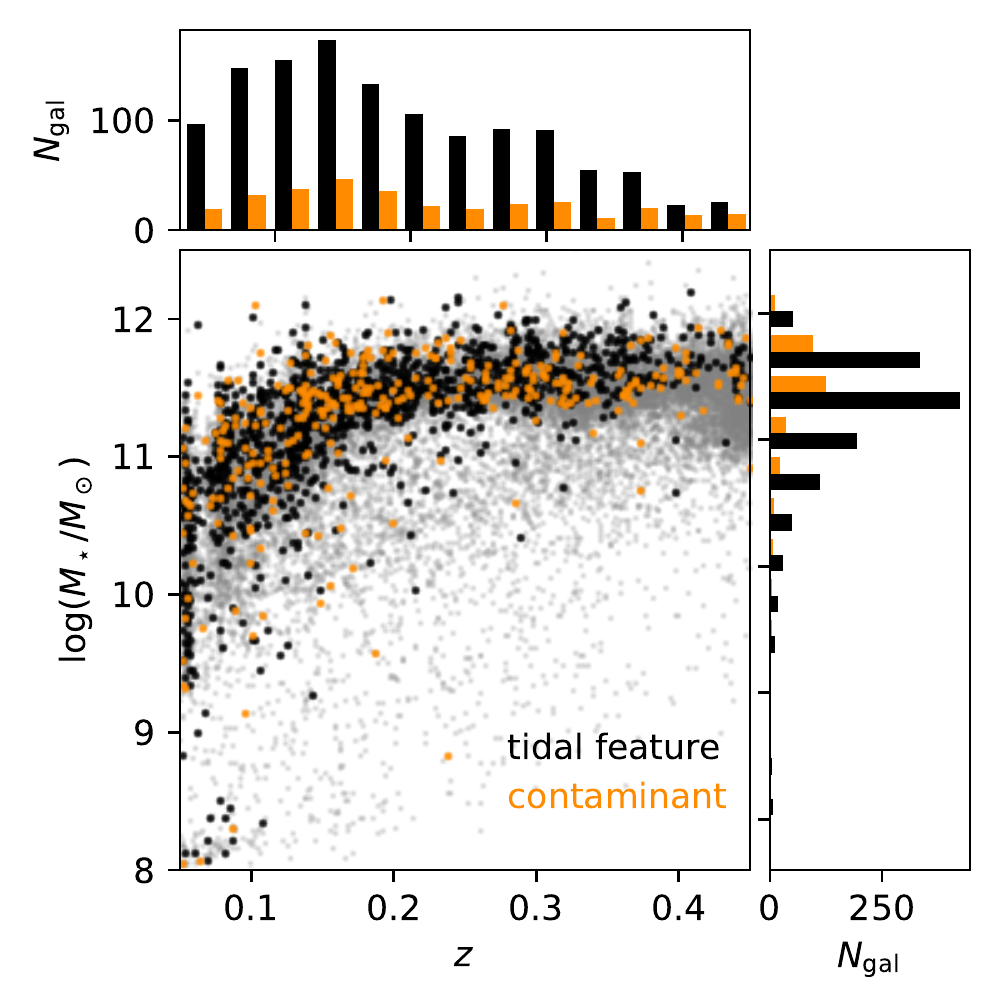}}
\caption{
Redshift v. stellar mass for the galaxies in parent spectroscopic sample (small grey points), the tidal feature sample (black) and the neighbor contaminant sample (orange). The one-dimensional histograms over redshift and stellar mass are shown in the top and right panels, respectively. In each case, the black bars show the distribution of the final, visually classified sample. In these panels, the orange bars show the neighbor contaminant sample, while the black bars show the tidal feature sample. The neighbor contaminants are not offset from the main tidal feature population, indicating that the visual removal does not introduce significant bias into the characteristics of our final tidal feature sample.
}
\label{vizcut}
\end{figure}

\begin{figure*}
\center{\includegraphics[width=\linewidth]{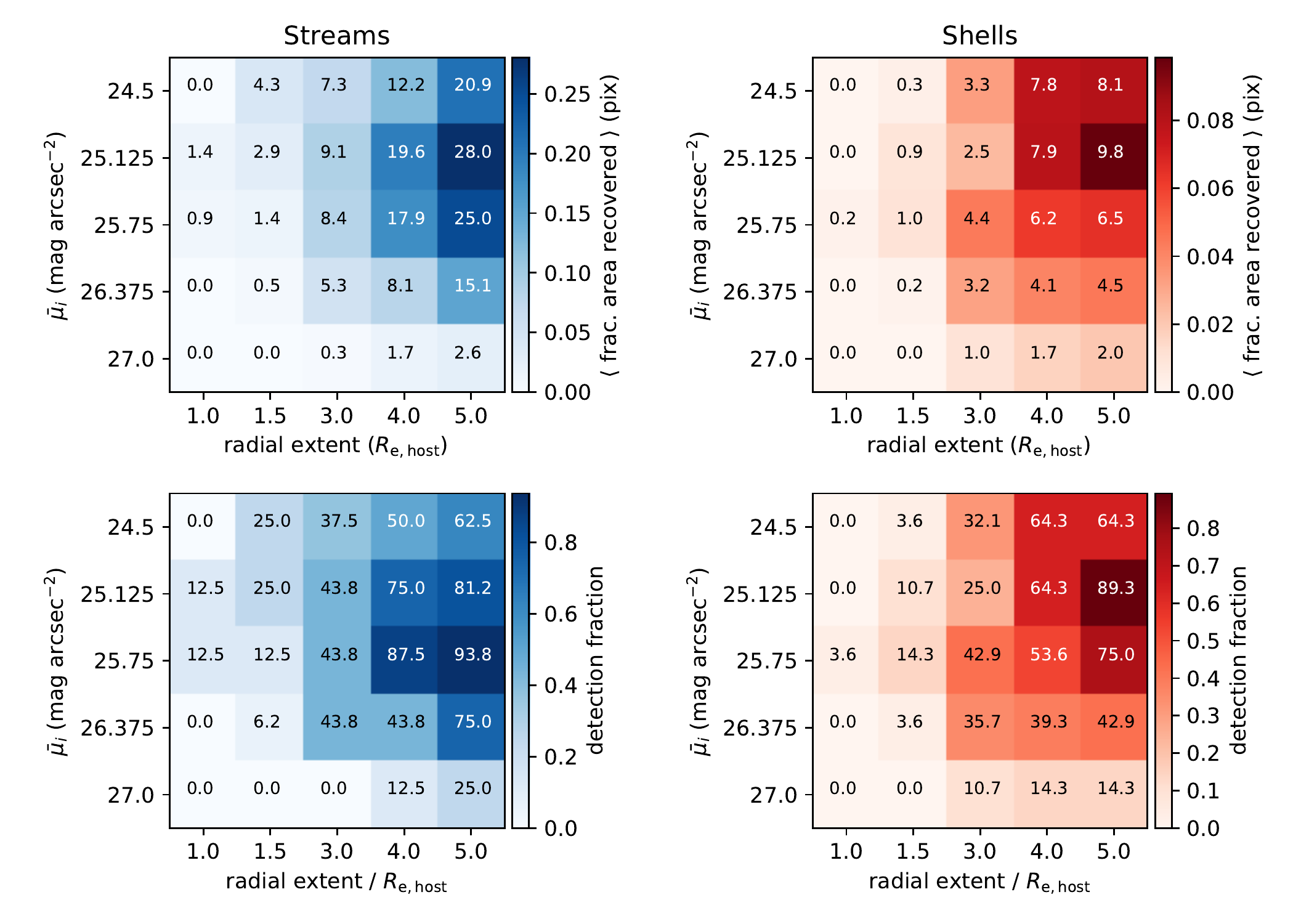}}
\caption{
\textit{Top row:} Pixel recovery fractions for simulated streams (left) and shells (right). Each box shows the fraction of pixels associated with the simulated tidal feature that are recovered by the algorithm for a given tidal feature mean \ihsc{} band surface brightness and radial extent, as measured in units of the host effective radius.  Generally, a higher fraction of pixels is recovered for the simulated streams; this is expected due to the number of particles occupying diffuse stellar fans in the shells. \textit{Bottom row:} Detection fractions for streams (left) and shells (right). Here, the values at each grid point show the percent of simulated tidal feature systems in which at least 5\% of the area covered by the tidal feature is flagged as a detection. In all panels, the drop in completeness at very high surface brightnesses ($\bar\mu_i \lesssim 25.$ \masq{}) is due to cases in which a part of the tidal feature is bright enough to be flagged as central galaxy substructure (e.g. spiral arms).
}
\label{pctrecovered}
\end{figure*}

In \autoref{detex}, we show the result of the algorithm on a synthetic case. Here, the signal which we wish to isolate is a simulated shell system from \citealt{hendel2015} with a mean surface brightness of $\bar \mu _i = 25.4$ \masq{}, where the mean is taken over pixels which have at least one count after the simulation is convolved with the HSC PSF (see \autoref{sect_detrecovery}). We add this signal to an \ihsc{} band image of a non-interacting galaxy from the parent sample (left panel). The detection image, $\mathbf{C}$, that we derive for the raw image, $\mathbf{I_0}$, is shown in the middle panel; the two bright caustics in the system are visible in the middle of the detection image. The detected tidal features output by the algorithm are shown by the orange outline in the rightmost panel.

In total, we detect candidate tidal features around \ntf{} of the galaxies in our sample. For this initial application, we take the additional step of visually classifying the morphology of the tidal features in the automated sample, as well as removing contamination from tidal features that are likely associated with projected neighbors.

\subsection{Visual Morphological Classification}
The decision to visually classify tidal feature morphologies and remove neighbors for this sample is driven by two factors. First, in order to build a classification algorithm which can differentiate shells from streams, it is necessary to have a sufficiently representative sample of known shells and streams from which an algorithm can be trained. There do exist simulated tidal features from which mock observations could be created for training \citep[e.g. ][]{kawata2006,johnston2008,hendel2015,amorisco2017,pop2017a} and tidal feature catalogs that overlap with the HSC Wide footprint \citep[e.g. ][]{nair2010,atkinson2013}. However, training solely from simulations would pose a significant danger of propagating biases due to the parameter space explored by the simulation.  The latter approach, training from known tidal feature hosts, is likely to yield large numbers of false negatives in the training set where HSC-SSP data reveal tidal features around a galaxy that was not identified previously, as HSC-SSP data reaches significantly lower surface brightnesses than previous surveys at similar areas. A training set from only those works that reach significantly lower surface brightnesses than HSC-SSP \citep[e.g.][]{vandokkum2005,martinezdelgado2009,tal2009} would have neither the necessary size nor be representative of the distribution of host galaxies in the sample at hand. 

Second, identifying the host of a tidal feature is in some cases a non-trivial problem \citep[see, e.g., ][]{greco2018}. For nearby galaxies with extended tidal features, the host galaxy is often not the source that is closest to the tidal feature. Future work will address automatic classification of such cases; for the sample at hand, we identify these cases visually. 

We visually classify each galaxy in the automated sample using four possible classes: stream, shell, ambiguous, and neighbor contaminant. Shells are characterized by their caustics, wherein the center of curvature is located near the center of the host galaxy. Streams are long, thin features wherein the curvature is not globally oriented towards the center of the host galaxy (though streams are not necessarily radial features). A small number of artifacts from HSC also passed our automatic cuts, and were removed at this stage. Amorphous features would comprise of all other features that do not fall into one of these categories; in practice, however, only one tidal feature system was classified by both human classifiers as amorphous, and so we neglect further analysis of this class. In cases where the detected tidal features are not visible to the human classifier, our visual classification is primarily governed by the shape of the detected regions. This produces a final sample of  \nshells{} shell hosts and \nstreams{} stream hosts between $0.05 < z < 0.45$.

In \autoref{vizcut}, we compare the redshift versus stellar mass distribution of our visually classified tidal feature sample (black) and visually identified contaminant sample (orange). We note that there is no apparent bias in the distribution of the visually classified neighbor contaminants with respect to our final sample.

\subsection{Detection recovery efficiency}\label{sect_detrecovery}
In previous work, the surface brightness limits of tidal feature searches have normally been estimated from the variations in fluxes from random sky apertures \citep{atkinson2013,hood2018,morales2018}. Using the methods described in \citealt{atkinson2013} and \citealt{morales2018}, we derive nominal surface brightness limits for the Wide layer of HSC-SSP of $\mu_i \ssim 27.9$ \masq{} and $\mu_i \ssim 28.4$ \masq{}, respectively. 

However, the detection of tidal features is also strongly dependent on the on-sky size and morphology of the tidal feature. We find that the surface brightness limits derived from the methods from prior work generally produce surface brightness limits that are significantly deeper than practical limits for both automatic and visual inspection (for more details, see \autoref{sect_sblim}). Moreover, our automated algorithm allows us to run through the detection process many times without a proportionate increase in the required human time. Thus, in order to understand the surface brightness and angular size limits of our detection method as applied to the HSC wide layer, we use the simulated tidal features of \cite{hendel2015} to create mock HSC-like tidal feature systems. 

Taking \ihsc{} band cutouts of six galaxies (4 elliptical, and 2 disc galaxies) in which no candidate tidal features were originally detected, we insert nine mock tidal features into each galaxy. We inject all tidal features such that the line of sight runs perpendicular to the orbital plane; the recovery fractions should therefore be taken as upper limits with respect to tidal feature orientation. The tidal features are scaled to a grid of mean surface brightnesses and sizes (normalized to the effective radius of the host) over $24.5<\bar\mu_i<27.0$ \masq and $1<d/R_e<5$, and convolved with the local HSC PSF. Here, $d$ refers to the maximum radial extent of the tidal feature, and $\bar\mu_i$ refers to the mean surface brightness of the convolved tidal feature system, averaged over the pixels in the cutout which have at least one count after convolution with the HSC PSF. Simulations of four streams and five shells are used for this test. 

For each simulated system, we define the recovery fraction as:

\newcommand{\frec}[0]{$f_{\rm rec}$}
\begin{equation}
f_{\rm rec} = \frac{N_{\rm pix, rec}}{N_{\rm pix,sim}},
\end{equation}

where $N_{\rm pix, rec}$ is the total number of pixels flagged as part of a detected tidal feature, and $N_{\rm pix, sim}$ is the total number of pixels in the PSF-convolved simulation that have at least one count. 

The results of these tests are shown in \autoref{pctrecovered}. The lefthand column (blue) shows the results for the simulated stream systems, while the righthand column (red) shows the results for the simulated shell systems. The top row shows, for a given radial extent and mean surface brightness, the average \frec{} of the simulated systems. The lower row shows, again for a given radial extent and mean surface brightness, the fraction of systems wherein \frec{}$>0.05$ (approximately the threshold at which a visual morphological classification can be made).

As expected, tidal features which extend farther from the center of the host galaxy are more easily detectable due to reduced host light contamination. For a given surface brightness and tidal feature extent, the simulated streams have a higher \frec{} than shells (upper row) and are more likely to be detected at visually classifiable levels than shells (lower row); this is a product of the fact that a significant number of the particles in the simulated shells are contained within diffuse stellar fans (for reference, see the low surface brightness fans in the right panel of \autoref{detex}). 

There is also a drop in detection recovery at very high mean surface brightnesses -- this occurs when the tidal feature enters a regime in which its associated signal in the detection map become comparable to that of spiral arms and neighboring sources. These tidal features are therefore removed as a likely contaminant (i.e. misidentified as a neighboring source or spiral arm). However, such features are in practice typically associated with ongoing major mergers, are are thus not the focus of this work. 

As shown in \autoref{pctrecovered}, the algorithm can identify tidal features down to $\approx 27$ mag/arcsec$^2$ at an extent of $d\gtrsim4R_e$. We note, however, that we are significantly incomplete at these surface brightnesses (for example, although we detect, on average, 2.6\% of the pixels of our simulated streams at $27$ mag/arcsec$^2$ with an extent of $d=5R_e$, only 25\% of simulated cases yield a detection). Our approximate 50\% completeness for tidal features with $d\geq3R_e$ is approximately $\bar\mu_i \approx 26.4$ mag arcsec$^{-2}$. 


\subsection {Measuing tidal feature color}\label{sect_tfcolordet}
For a subset of the galaxies with detected tidal features, the features are sufficiently far from the galaxy to allow for a clean measurement of color. We rerun our detection algorithm in the \ghsc{} band, and measure the color of the detected tidal features by using the union of the detected regions in \ghsc{} and \ihsc{} as the apertures (for example, in \autoref{detex} the orange outlines in the rightmost panel would be used as an aperture).

To determine where the color of the tidal feature will be significantly impacted by flux from the host galaxy, we measure the tidal feature contrast ratio, \ctf{}, as the observed flux of the system and expected flux from the host via surface brightness profiles where the tidal features have been masked from the image, i.e. $\mctf{}=f_{\rm observed}/\hat f_{\rm host}$. We measure $f_{\rm observed}$ by simple aperture photometry, where the aperture is the detected region associated with the tidal feature. We estimate $\hat f_{\rm host}$ by measuring the surface brightness profiles of our sample in the \ghsc{} and \ihsc{} bands in elliptical annuli. Here, the position angle and ellipticity are obtained from fits to the host galaxy using \textsf{imfit} \citep{erwin2015}, where neighboring galaxies and the tidal feature system are masked from the fit. $\hat f_{\rm host}$ is then the flux expected in the region of the tidal feature as estimated by the surface brightness profile in that region. 

To calibrate and test the ability of \ctf{} to trace the accuracy in measurements of colors for extended tidal features, we inject tidal feature simulations (presented in \autoref{sect_detrecovery}) of known colors into non-interacting HSC galaxies. In these simulations, the particle data are also convolved to the HSC PSFs in each band. The simulated systems are then run through our full detection and color measurement algorithm. We find that a \ctf{} threshold of 2.5 produces tidal feature colors which are accurate to 0.06 mag, which includes the effect from band-dependent PSF size.

The color of the candidate tidal feature with respect to the host also acts as a flag for potential neighbor contamination. When comparing tidal feature color to host color, we consider \gi{}, where \gi{}  is defined as the difference between the $(g-i)$ color of the tidal feature and the host (i.e. larger values correspond to a tidal feature that is redder with respect to its host). In the case of shells, we keep only those tidal feature candidates with $-0.5<$\gi{}$<0.2$ mag. For this initial sample, we verify visually that this color cut does not remove potentially interesting tidal features from the sample. Streams have been observed to host significant star formation \citep{knierman2012, higdon2014}, and so we impose only the red cut, i.e. we keep those color measurements for which \gi{}$<0.2$ mag. 

After removing likely neighbors, we estimate an average color of the tidal feature system. We estimate the uncertainty on the measurement of the average color ($\sigma_{\overline {\rm TF} }$) as follows:

\begin{equation}
\sigma_{\overline {\rm TF}} = (\sum_i^N \sigma_{{\rm TF}, i}^2 + \sigma_{\rm sim})^{1/2},
\end{equation} 

\noindent{}where $ \sigma_{{\rm TF}, i} $ is the uncertainty on the color of the $i^{\rm th}$ detected tidal feature region (as propagated from the variance maps output by \textsf{hscPipe}) in a system of N tidal feature regions (defined as a group of contiguous pixels that are flagged by the detection algorithm; for example, \autoref{detex} shows a $N=2$ system), and $\sigma_{\rm sim}$ is the uncertainty due to PSF differences and host galaxy contamination ($\sigma_{\rm sim}\ssim 0.06$, from above). 

Finally, we estimate the color of the host galaxy core from simple aperture photometry with an aperture diameter of $2R_e$. We do not perform PSF matching for this measurement, as the impact of doing so is negligible. 

\begin{figure*}
\center{\includegraphics[width=0.8\linewidth]{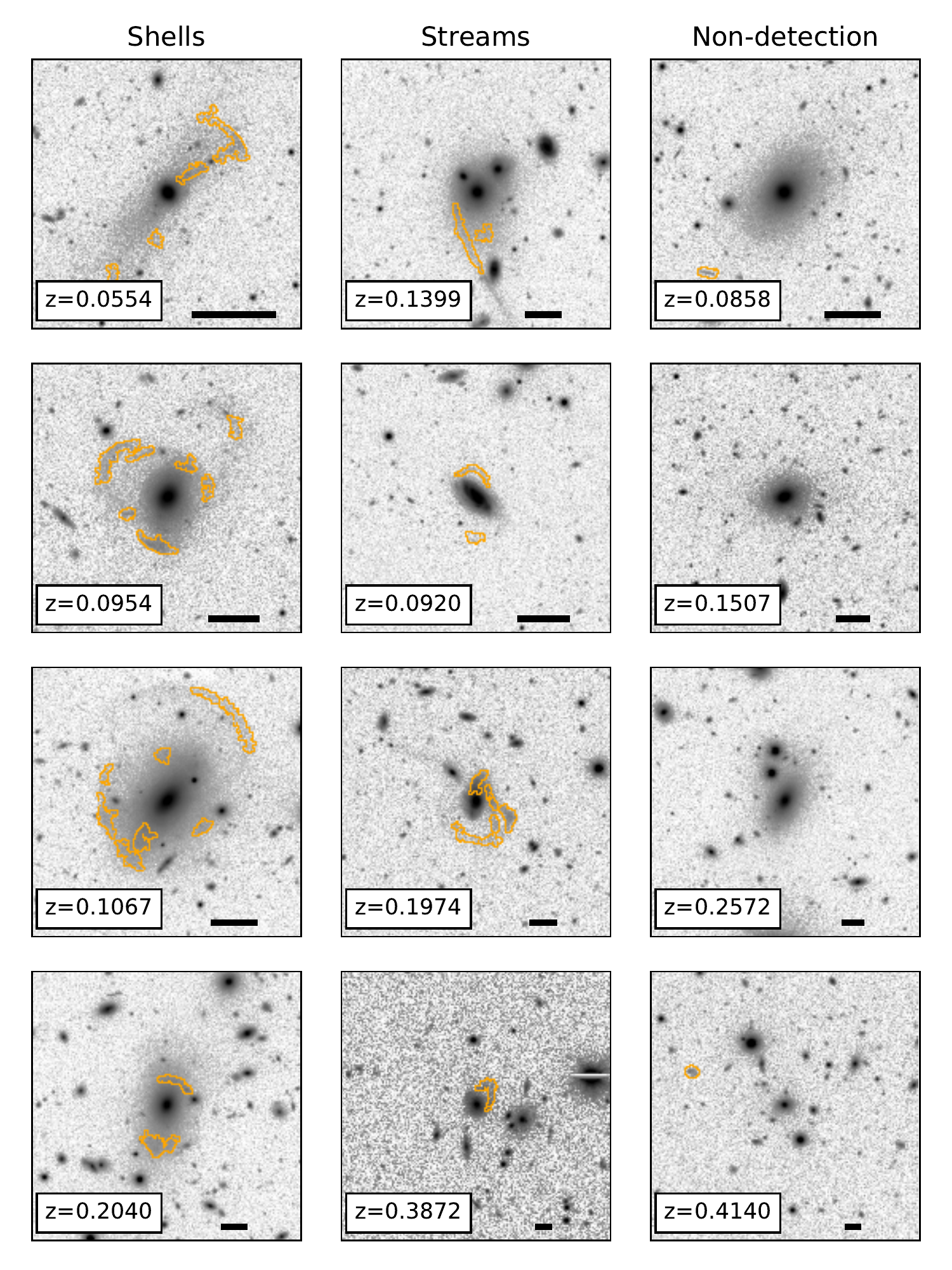}}
\caption{
A sample of \ihsc{} band imaging of shells (left column), streams (middle column) and non-interacting galaxies (right) in our sample. The orange outlines show the boundaries of the detected regions found by our filtering algorithm. 
The horizontal black line in the bottom right of each panel spans 30 kpc at the redshift of the central galaxy. In all panels, a logarithmic stretch is used to emphasize low surface brightness features.
}
\label{showsample}
\end{figure*}

\begin{figure}
\center{\includegraphics[width=\linewidth]{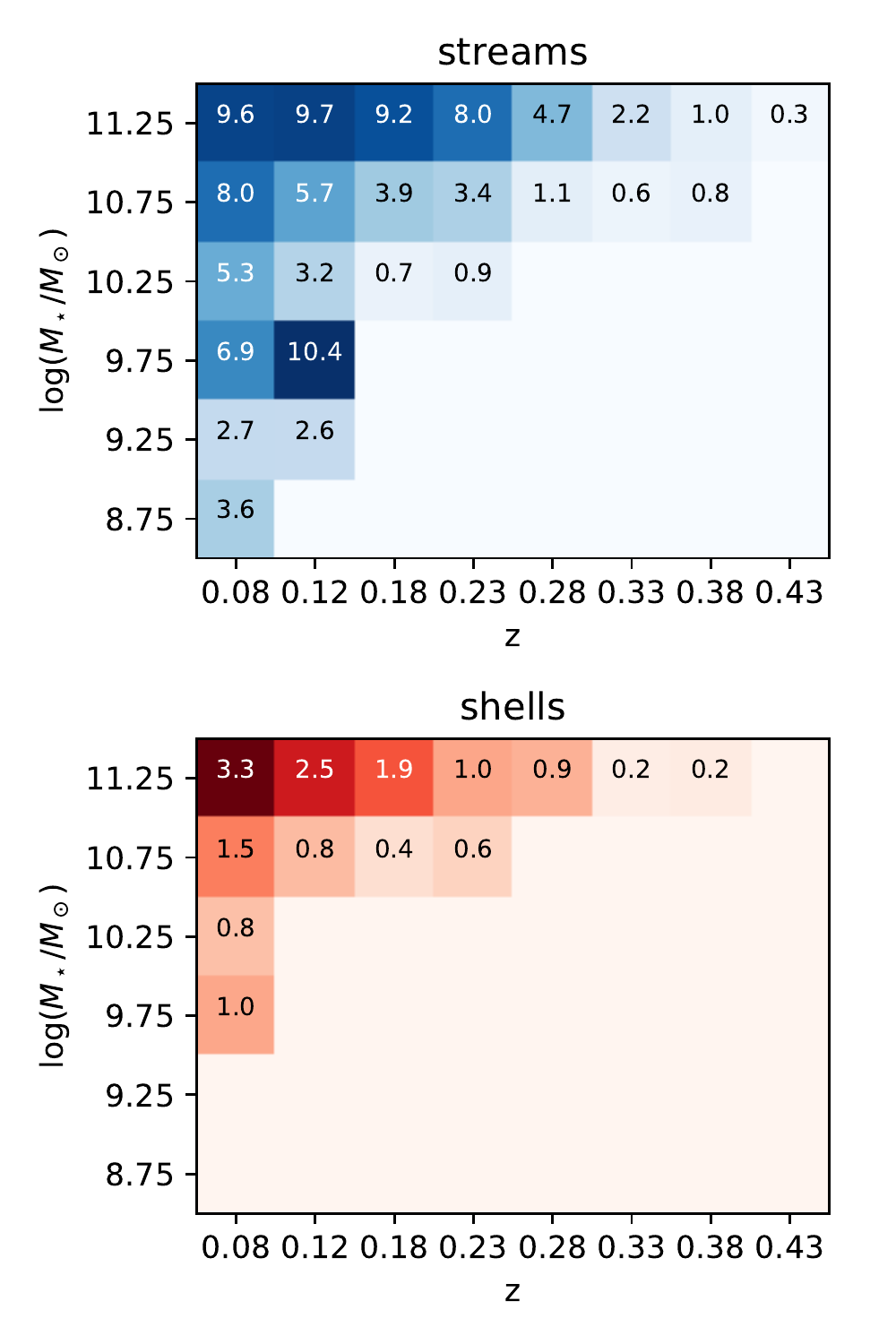}}
\caption{
Tidal feature occurrence fractions as a function of stellar mass (y-axis) and redshift (x-axis) for the visually identified stream (top) and shell (bottom) samples. Each bin in the top (bottom) panel is annotated with the fraction of galaxies tagged as a stream (shell) host from the full spectroscopic sample. Blank regions indicate bins for which no tidal feature systems were identified.
}
\label{occurrencefractions}
\end{figure}

\section{Results}\label{sect_results}
Here, we present a sample of \nshells{} shells and \nstreams{} streams found semi-automatically in the S16A internal release of the HSC-SSP Wide layer (covering $\ssim200$ deg$^2$). The tidal feature morphologies are visually classified as either streams or shells, and cover a redshift range of $0.05<z<0.45$. In \autoref{showsample}, we show a sample of shell (left) and stream (middle) galaxies, and galaxies without tidal features (right) as identified by this method.

\subsection {Observed tidal feature occurrence fractions}\label{sect_occfrac}
We first consider the tidal feature occurrence fractions for our sample, which we define to be the fraction of galaxies in the sample within a given range of redshift and stellar mass in which a stream or shell was detected, divided by the total number of galaxies in the sample that occupy that area of parameter space. As shown in \autoref{sect_detrecovery}, the observed occurrence fraction will reflect both the astrophysical occurrence fraction and our ability to recover the tidal feature signal from HSC data.

To first order, for a typical tidal feature in our sample (i.e. averaging over time since interaction, orbital parameters), the surface brightness and angular size of the tidal feature should be primarily dependent on the redshift and the stellar mass of the source. \autoref{occurrencefractions} shows our observed occurrence fractions as a function of host redshift and stellar mass. As expected, our observed occurrence fractions fall as redshift is increased or stellar mass is decreased. The observed occurrence fractions of shells falls more steeply as a function of redshift than than that of streams, which we expect from our lower recovery fractions for simulated shells in \autoref{pctrecovered}. At low redshift ($z<0.15$), we find a number of low mass systems with $\smass < 9.7$ that are apparent stream hosts. We find that these systems are either dwarf galaxies undergoing major mergers (due to the mass of the galaxies involved, the surface brightness of the extended features is low enough to avoid being thrown out as spiral arms), or close neighbors of larger galaxies that host tidal features. Though ongoing major mergers in dwarf galaxies are not the focus of this paper, we note that our method could also generate samples of such ongoing mergers if a different parent sample were used.

The fraction of galaxies which host observable tidal features has been measured by many authors who find a wide range of values, from 7\% as derived from SDSS \citep{nair2010} to 71\% from a deep imaging campaign of local ETGs \citep{vandokkum2005}. It is likely that a majority of faint substructure have peak surface brightnesses lower than $30$ \masq{} \citep{johnston2008}, and that the occurrence and morphology of tidal features is dependent on the mass and merging history of the host \citep{wang2005, jiang2015, hendel2015}. Thus, tidal feature detection is highly sensitive to the type of host galaxy examined and to the surface brightness limits of the relevant observations, leading to a large amount of variance into observed tidal feature occurrence fractions. 

In campaigns where a smaller area on the sky is observed to large depths, the limiting surface brightness is significantly deeper than that which is accessible from the HSC wide layer \citep[e.g. ][]{vandokkum2005,tal2009,martinezdelgado2010}. Though we are unable to construct a matching sample against which to compare our observations, we note that the systematically higher occurrence fractions that are often found by such studies point to the minor merger picture, in which increasingly more numerous and minor mergers leave behind increasingly fainter tidal features \citep{johnston2008}. 

Other studies present samples which we have no clean method of emulating. For example, \cite{schweizer1988} considers S0, S0/Sa, and Sa NGC galaxies and derives an occurrence fraction of $16\%$, while \cite{adams2012} observes a $3\%$ tidal feature occurrence fraction for cluster ETGs. 

However, it is valuable to consider whether the tidal feature occurrence rate that we observe from our semi-automatic method is equivalent to those samples derived from visual classification alone. Here, we compare our observed occurrence fractions to literature results where it is possible to account for differences in sample selection.

\subsubsection{Comparison to \cite{atkinson2013}}
The catalog of tidal feature hosts in the Canada-France-Hawaii Telescope Legacy Survey (hereafter CFHTLS) assembled by \cite{atkinson2013} is perhaps the clearest benchmark against which we can evaluate our observed tidal feature occurrence fractions. The data of CFHTLS are similar to the HSC Wide layer in terms of depth and spatial resolution; the deepest band that \cite{atkinson2013} considers, the CFHT $g$ band, has a roughly equivalent $5\sigma$ point source depth as \ihsc{} in the Wide layer of HSC-SSP \citep{boulade2003, aihara2018}. The catalog was constructed entirely from visual inspection, and thus serves to test the efficiency of our method against this approach.

To compare our tidal feature detection fractions to those of \cite{atkinson2013}, we consider the 2113 galaxies in our sample for which $15.5<r_{\rm SDSS}<17$ mag, $M_r <-19.3$ \citep[derived from SDSS, ][]{alam2015} and $0.05<z<0.2$, following the sample selection cuts used by \cite{atkinson2013}. We find a tidal feature occurrence fraction of $13.30 \pm 0.79\%$, where the uncertainty reflects only counting uncertainty assuming a Poisson distribution (this value is significantly higher than those in \autoref{occurrencefractions} because of the additional cuts for bright galaxies imposed in the \citealt{atkinson2013} sample).

Within the \cite{atkinson2013} sample (hereafter the CFHTLS sample), we consider only the galaxies for which the authors were able to make a morphological classification of the tidal feature. We furthermore exclude any galaxies for which the only feature morphology tags belong to the classes ``diffuse'' and ``fan'', which are observed in $4.6\%$ of their galaxies, as the nature of our method biases us significantly against the detection of tidal features which are diffuse (i.e. of similar to or lower spatial frequency to their host galaxy).  

Given these cuts, the fraction of galaxies that host morphologically classifiable tidal features of \cite{atkinson2013} is, at $12.98 \pm 0.85\%$, in agreement with our observed tidal feature occurrence fraction. 

Given that the \cite{atkinson2013} was constructed via purely visual classification from a dataset with a similar limiting depth and spatial resolution as HSC-SSP, this agreement is encouraging, and implies that our detection method is able to detect tidal features as well as visual inspection methods in the regime of shell-like and stream-like tidal features.

\subsubsection{Comparison to \cite{malin1983}}
As one of the earliest papers to report the occurrence rates of shells around elliptical galaxies, it is informative if not entirely straightforward to compare our shell occurrence fraction to that of \cite{malin1983}. The authors consider a sample of NGC ellipticals from the ESO/SRC (IIIa-J) Southern Sky Survey, and find that 5.8\% of the galaxies therein show signs of stellar shells. The authors cite their surface brightness limit as $B=26.5$ \masq{} on photographic plates.

Although the galaxies considered by \cite{malin1983} are significantly below our lower redshift range, we do not expect the true shell occurrence fraction to change significantly out to $z\lesssim 0.1$ \citep{pop2017a}. We use concentration as a proxy for morphology, wherein high-concentration galaxies are likely ellipticals, and low-concentration galaxies discs. We consider only high-concentration ($C>3.1$, where $C$ is defined as the ratio between the the radii containing 90\% and 50\% of the Petrosian i-band luminosity as measured in SDSS) galaxies at the lowest redshifts that we can probe ($0.05<z<0.1$). We set the cut in concentration following \cite{strateva2001} and \cite{dsouza2014}, but increase our threshold to obtain a more pure sample of ellipticals. This subset of our sample gives a shell occurrence fraction of $6.13\pm 1.23\%$, which is in agreement with that of the \cite{malin1983} sample.

\begin{figure*}
\center{\includegraphics[width=\linewidth]{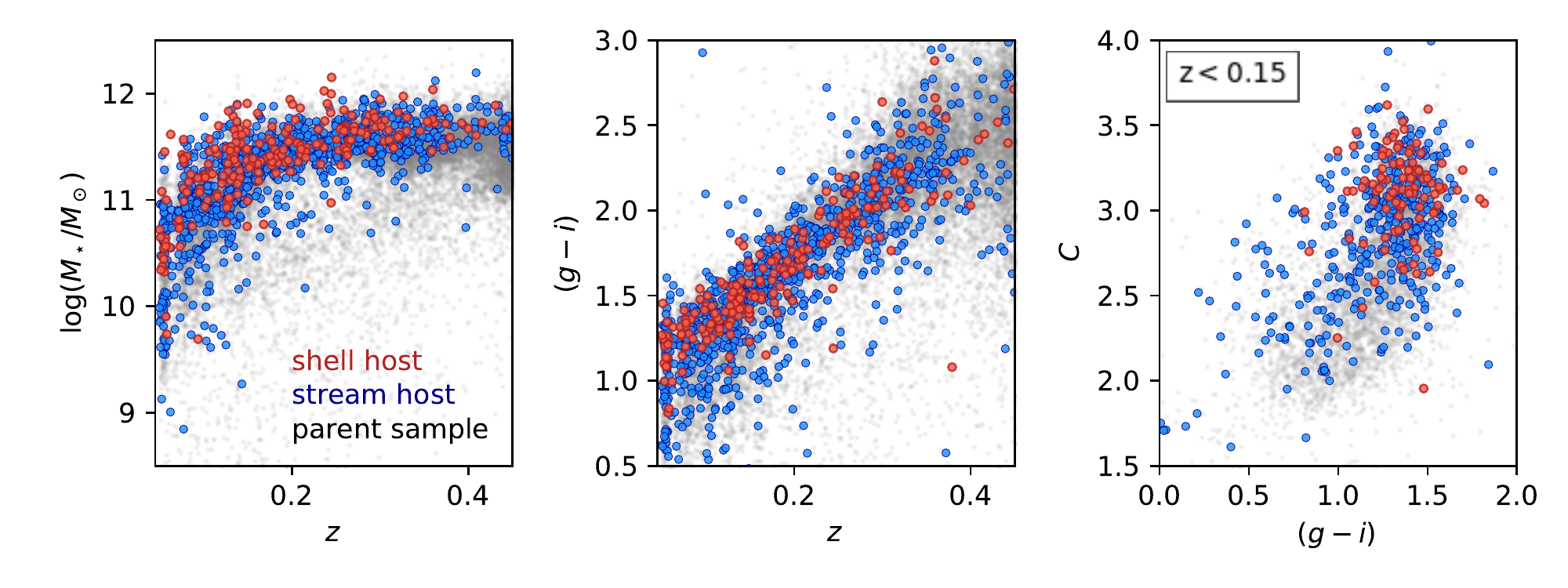}}
\caption{
Properties of the stream (blue) and shell (red) host galaxies. The small grey points show the full spectroscopic sample within our target redshift range. \textit{Left:} redshift versus stellar mass for the tidal feature sample.  \textit{Center:} redshift versus $(g-i)$ color for the sample. At all redshifts, shell hosts tend to be redder and more massive than their counterparts that host streams. \textit{Right:} $(g-i)$ color versus concentration for the $z<0.15$ galaxies in our sample. The shell host galaxies have, on average, higher concentration than the stream hosts, indicating that the shell sample contains a larger fraction of ETGs than does the stream host galaxy sample.
}
\label{hostprops}
\end{figure*}

\subsection{Host galaxy properties}

We first review the properties of the galaxies that host shells and streams in the context of the full SDSS spectroscopic sample. Relative to the non-interacting (i.e. galaxies for which no tidal features were detected) sample, shell hosts tend to have higher stellar masses and redder $(g-i)$ colors at a given redshift (left and middle panels, \autoref{hostprops}). Stream hosts span the full range of stellar masses and colors present in the non-interacting sample at a given redshift. 

\begin{figure*}[ht!]
\center{\includegraphics[width=\linewidth]{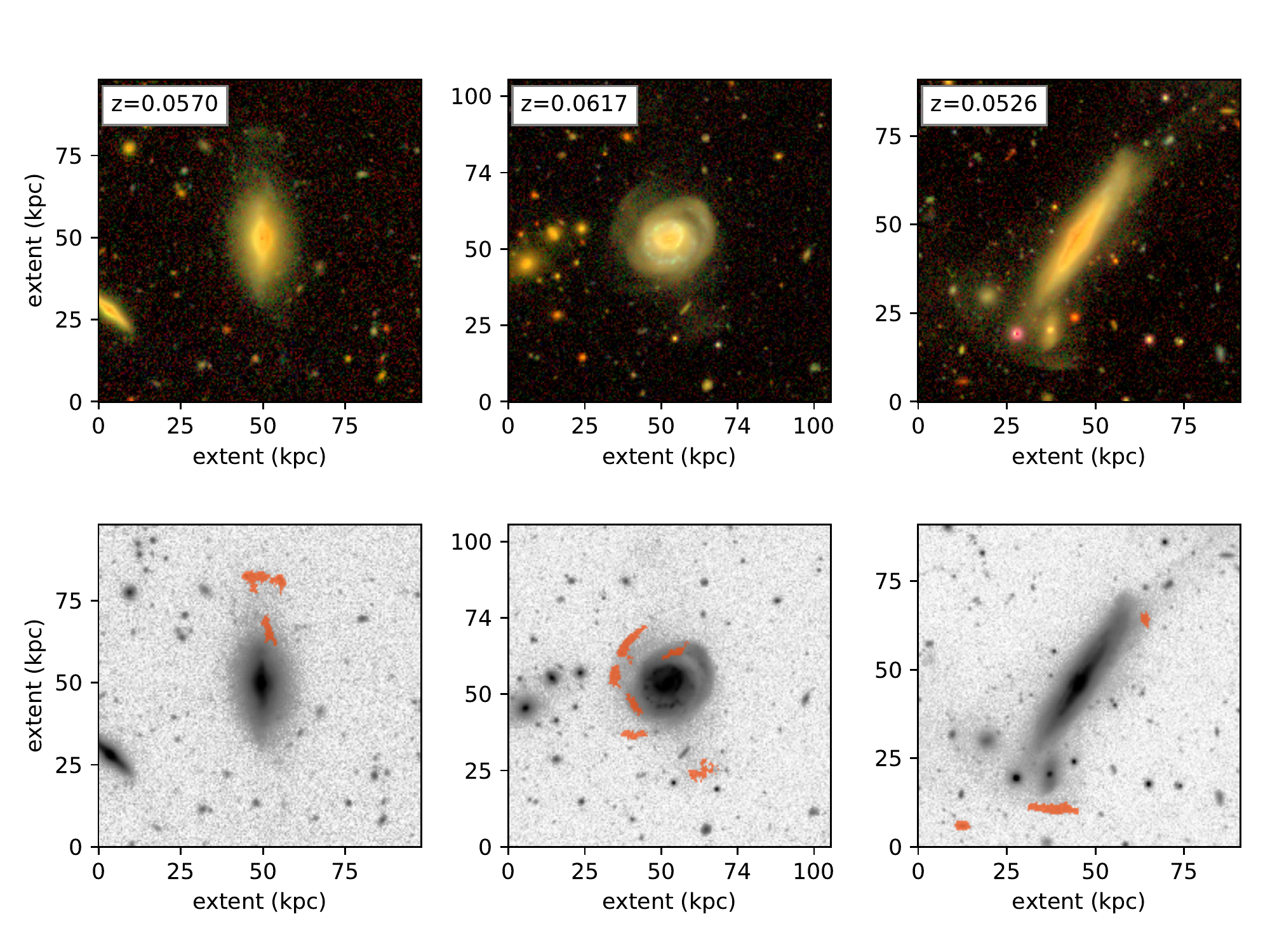}}
\caption{
\textit{Top:} $gri$-composite images \citep{lupton2004} of the three disc galaxies that host shells in our sample. Both caustics and diffuse stellar fans are visible in all cases. In the middle panel, two umbrella-like shell structures are visible to the left of the galaxy. \textit{Bottom:} The same galaxies, with the features detected by the algorithm shown in orange. A diffuse trail is also visible in to the top right of the rightmost panels.
}
\label{spiralshells}
\end{figure*}

At $z<0.15$, stream host galaxies in our sample have higher concentrations on average than do the non-interacting sample. The shell host galaxies have higher average concentrations than do either the stream hosts or the non-interacting sample (\autoref{hostprops}, right panel). These offsets imply that the shell hosts tend to be massive, red ETGs. From a visual inspection of our sample, we find only three instances in which a disc galaxy hosts a shell (see \autoref{spiralshells}). 

\begin{figure}
\center{\includegraphics[width=\linewidth]{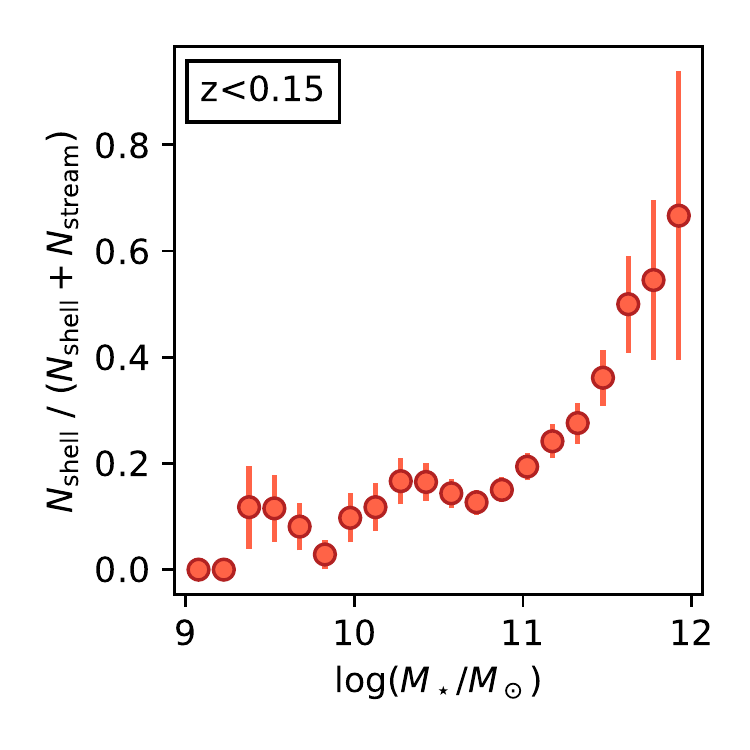}}
\caption{
Stellar mass versus the fraction of tidal features that are shells for the $z<0.15$ galaxies in our final tidal feature sample. Shells are found more frequently around more massive galaxies, growing from $\approx 20\%$ at $\log ( M_\star / M_\odot ) = 10.5$ to $\approx 50\%$ at $\log ( M_\star / M_\odot ) = 11.5$. A larger shell fraction around massive galaxies is expected due to an increase in the number of radially-biased satellite infall trajectories around such galaxies \citep{jiang2015,hendel2015}. 
}
\label{shellfrac}
\end{figure}

In a paradigm where satellites falling into more massive hosts tend to be on more radial infall trajectories \citep[see, e.g.][]{benson2005, jiang2015}, we expect somewhat fewer spiral galaxies to host shells due to the differences in the stellar mass range occupied by elliptical and spiral galaxies. To gauge whether the low number of shells observed around discs is purely a mass effect, we show the fraction of tidal features that are identified as shells in our sample as a function of stellar mass at $z<0.15$ in \autoref{shellfrac}.  

To estimate the effect of differential completeness in shell and stream detection on the trend apparent in \autoref{shellfrac}, we take the naive null hypothesis that shells and streams appear equally often, independent of stellar mass, and have radial extents that are distributed uniformly from 3$R_e$ to 5$R_e$. We add to this naive hypothesis the assumption that all the shells and streams are formed at the same time, that the events all have the same mass ratio, and that the surface brightness of the tidal feature for a given mass ratio drops linearly as a function of host stellar mass. In this case, using the outcomes of the simulations presented in \autoref{sect_detrecovery}, the fraction of observed shells would remain roughly constant as a function of host stellar mass, and rise at low masses. Thus, we expect that the trend seen in \autoref{shellfrac} is not a completeness effect.

If we take the naive assumption that the morphology of the tidal feature is independent of the morphology of the host, we would expect $\approx 15\%$ of detectable tidal features around low-concentration galaxies to be shell-like at $\smass{} \ssim{} 10.5$, near the peak of the distribution of stellar mass for a sample of low-concentration galaxies in our sample. However, we find that only $4\% \pm 3\%$ ($6\% \pm 4\%$) of the tidal features around $C<2.8$ ($C<2.6$) galaxies at $10.2<\smass{}<10.7$ are shell-like.  

We repeat this exercise using morphological classifications from the citizen science project Galaxy Zoo \citep{lintott2011}. Here, we consider only those galaxies where one class accounts for $>80\%$ of the votes. We again find that, at the same stellar mass, disc galaxies show a lower shell fraction than ellipticals, though the low number of shell systems around disc galaxies prevents us from making a statement about the disc galaxy shell fraction as a function of stellar mass.

We also consider the possibility that our detection method is biased against finding shells around face-on discs. To first order, contamination by spiral arms should affect the detection of both shells and streams, rather than the balance between the two. Using the method outlined in \autoref{sect_detrecovery} to compare the detection recovery efficiency of shell-like tidal features of a face-on spiral to that of an elliptical host with similar on-sky size does not show significant differences in detection recovery for the automatic stage. However, it is probable that human classifiers are biased against shell-like tidal features around face-on disc galaxies, as the curvature of shell caustics is similar to that of spiral arms. 

To test whether a human bias can account for the difference in shell fraction at fixed stellar mass, we consider only shells which are substantially separated from their host galaxy (i.e. high values of $c_{TF}$, see \autoref{sect_tfcolordet}), and find that we again observe relatively more ellipticals that host shell systems at fixed stellar mass. We therefore conclude that the relative dearth of shell systems around disc galaxies in our sample is not due to a bias in our sample selection method. 


We find no significant differences between the color gradients of the tidal feature host galaxies and a non-interacting sample matched in mass, redshift, and color -- this implies either that the average tidal feature is generated by a minor merger that does not significantly alter the color gradient of the host, or that the tidal feature was created sufficiently long ago so as to allow for relaxation. This is in agreement with the findings of \cite{kim2012}, who find that the bulk structural properties of tidal feature hosts are not significantly different than those of non-interacting galaxies.

\begin{figure}
\center{\includegraphics[width=\linewidth]{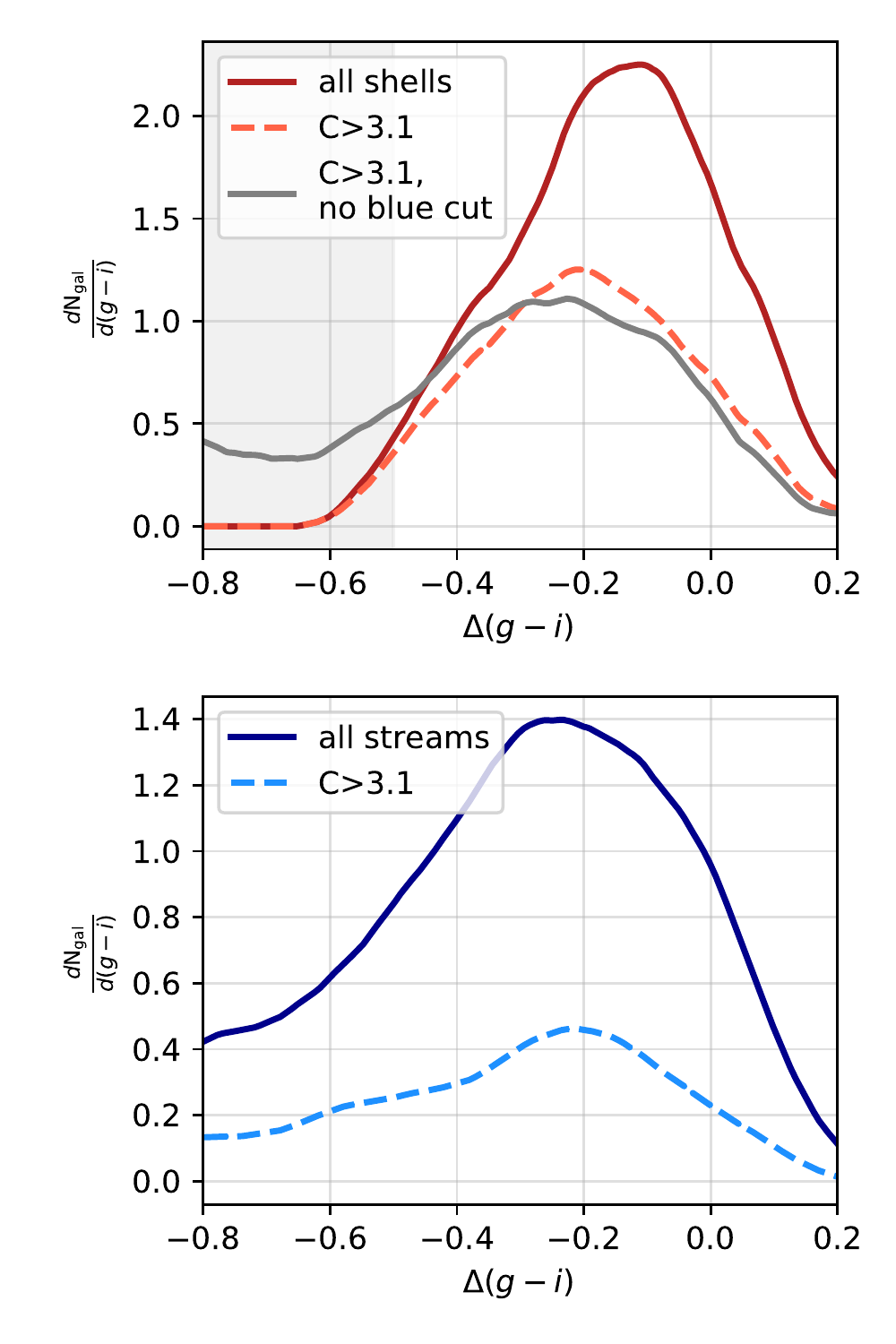}}
\caption{
Scaled kernel density estimates of the difference between the color of the tidal features and the core of the host galaxy (measured within a radius of $1R_e$) for the shells (top panel, red) and the streams (bottom panel, blue) via a kernel density estimate (using an Epanechnikov kernel with a width of $\Delta(g-i)=0.15$). In each panel, the solid lines show the color distribution of the full sample. The dashed lines show the \gi{} distribution for our high-concentration ETG sample ($C>3.1$), normalized to the number of high-concentration galaxies in the sample. In the top panel, the grey solid line shows the distribution of \gi{} for the shells that we would have derived if we did not remove features with measured \gi{}$<-0.5$ as neighbors. The grey shaded region in the top panel shows the color cut that we implement on the color measurements of individual tidal features to remove neighbors in the shell sample. 
}
\label{colordists}
\end{figure}

\subsection{Tidal feature colors}\label{sect_colormeas}
We are able to measure colors for \ncolshells{} shell systems and \ncolstreams{} stream systems in our sample. Below, we present the average $(g-i)$ colors of tidal feature systems. Here, we specifically consider the difference in the average color of the tidal feature system and the core of its host galaxy (hereafter \gi{}). In the case of a non-star-forming satellite, \gi{} is a probe of the mass ratio of the merging event that formed the shells: if one assumes that the tidal feature progenitor satellites sat on the red sequence during their infall, the mass of the satellite may be estimated by finding the mass on the red sequence that corresponds with the measured satellite color \citep{gu2013}. We find this assumption to be reasonable, as the red colors measured for the shells are too red to plausibly host significant star formation. In the case of blue tidal features, \gi{} does not yield mass information.

\subsubsection{Shell colors}

We find that the shells are, on average, slightly bluer than their hosts, with a mean color offset of $-0.15 \pm 0.02$  mag (see top panel of \autoref{colordists}). The observed \gi{} does not have a strong dependence on the mass of the host galaxy. This finding is in agreement with the majority of individual systems in the literature for which the color of the shells have been measured \citep{quinn1983,gu2013,foster2014}.  The color difference between the shells and their hosts then suggests that our sample is dominated by mergers where the infalling satellite is significantly less massive than the host. 

However, we also find that for $15\% \pm 4.4\%$ of shell galaxies, the shells are consistent, within errors, with being the same color as the core of their host galaxy. We will hereafter refer to these galaxies as \redshells{}. This lack of a color difference has been suggested by \cite{pop2017b} to be a signature of shells produced by major mergers, wherein the shell caustics of the system within $\approx 50$ kpc from the host center approach the metallicity to the core of the host galaxy. Though the shell caustics in \cite{pop2017b} have slightly lower metallicities than the host core, we estimate that their optical colors will be consistent within error to the core of the host within $1 R_e$.

\begin{figure*}
\center{\includegraphics[width=0.6\linewidth]{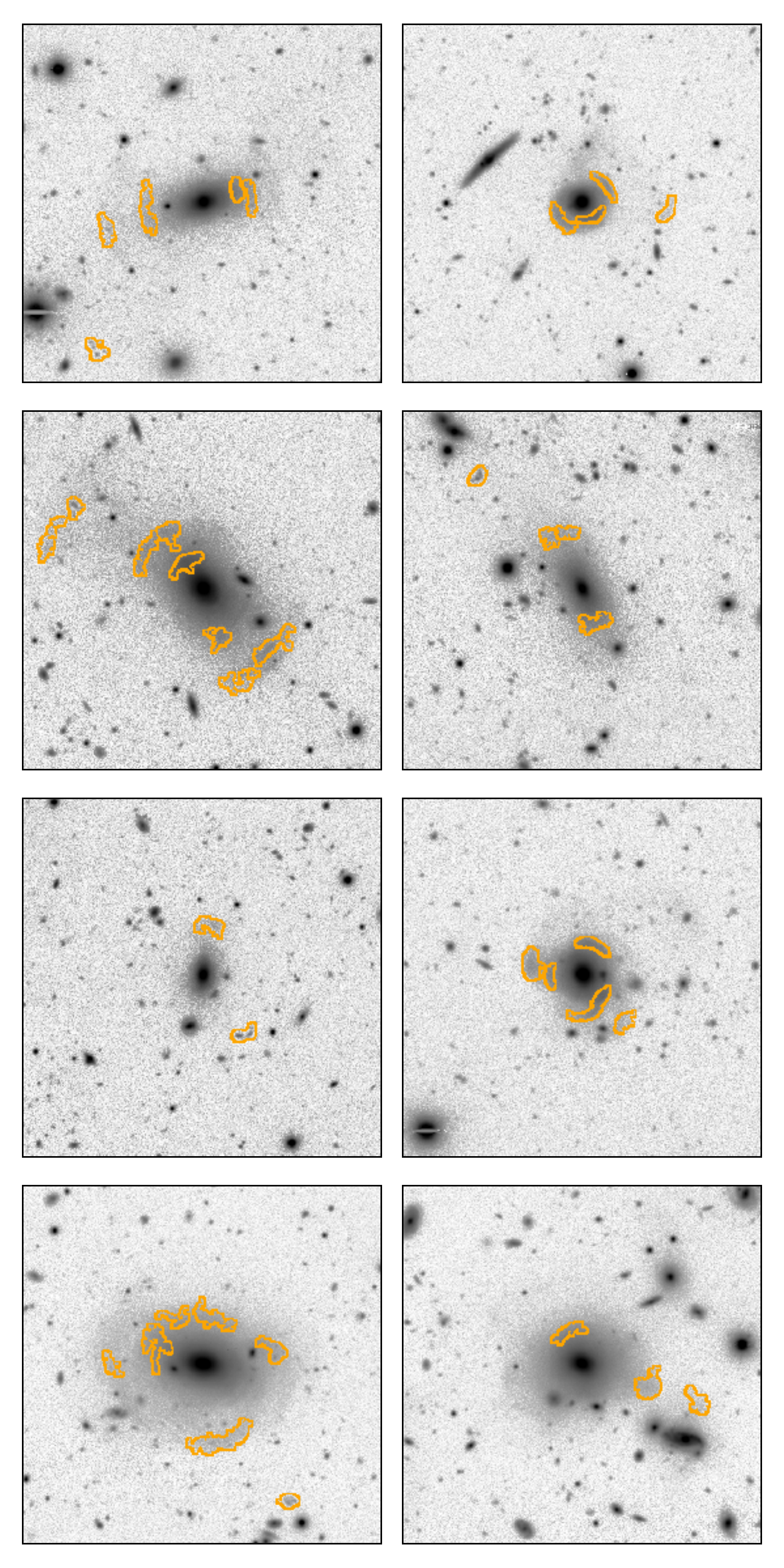}}
\caption{
\textit{Left:} \ihsc{} band imaging of a sample of shells where the color of the shell caustics (boundaries defined by the orange outlines) is consistent with the color of the host galaxy within $1R_e$. \textit{Right:} a sample of shells for which the color of the shell caustics is bluer ($\Delta m<-0.25$ mag) than the color of the host galaxy within $1R_e$. All panels use a logarithmic stretch to increase contrast at low surface brightness. The shells that are as red as their host galaxy cores display a higher incidence of type I shells ($86 \pm 35\%$) than their blue counterparts ($30 \pm 14\%$). 
}
\label{popshells}
\end{figure*}

Though there are relatively few \redshells{}, we find there to be a morphological difference in these shells. Relative to the other galaxies that host shells, these \redshells{} tend to host more Type I shells, wherein the axis of symmetry of the shells is aligned with the major axis of the galaxies (see the leftmost panel of the second row in \autoref{popshells} for an example). The caustics of Type II shells, on the other hand, are arranged randomly around the galaxy (see the middle panel of the bottom row in \autoref{popshells} for an example). The Type I shell occurrence fraction in our ``red shell'' sample is $86 _{-35}^{+14}\%$, significantly higher than the Type I shell occurrence fraction for the rest of the shell hosts,  $30\pm14\%$. Considering both the \redshells{} and minor merger-like shell hosts together gives a Type I host fraction of $43 \pm 12\%$. This is in agreement with the same quantity estimated by \cite{prieur1990} for the catalog of \cite{malin1983}. The uncertainties listed are those derived from counting statistics only, and reflect the significant uncertainty that we face due to low numbers of \redshells{} present in the sample.

\begin{figure}
\center{\includegraphics[width=\linewidth]{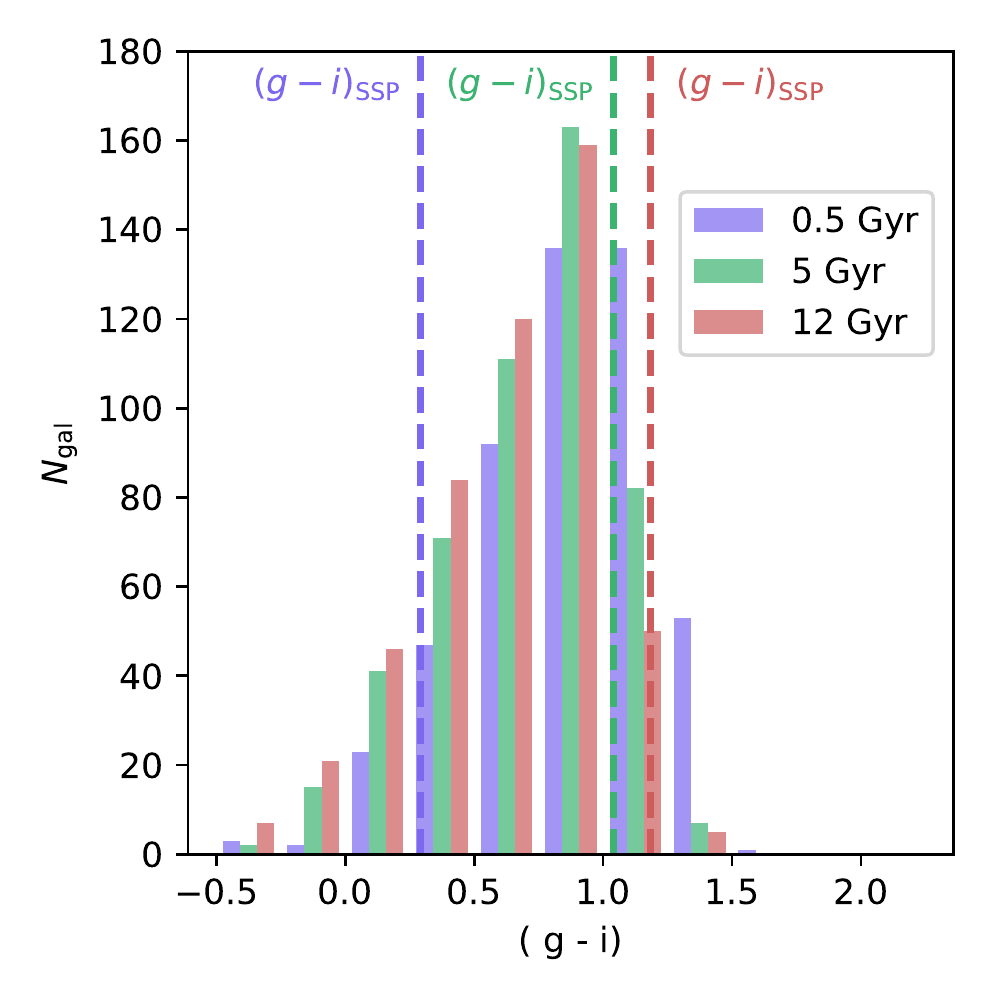}}
\caption{
The distribution of $(g -i)$ restframe colors for our stream sample, using the K-correction computed for a 0.5 Gyr (purple), 5 Gyr (green), and 12 Gyr (red) SSP from \citealt{bruzual2003}. The dashed vertical lines show the restframe colors for, from left to right, the 0.5 Gyr (purple), 5 Gyr (green) and 12 Gyr (red) populations. The existence of streams with rest-frame colors bluer than the youngest SSP considered implies that a subset of the streams host star formation.
}
\label{kcorrstreamcolors}
\end{figure}

\subsubsection{Stream colors}
The streams in our sample are bluer with respect to their hosts than the shells are, with a mean offset of $-0.38 \pm 0.02$ mag (see bottom row of \autoref{colordists}).  There are also a number of very blue streams around less massive galaxies. We bound the restframe colors of these streams by calculating K-corrections for three SSPs from \citealt{bruzual2003} with a \cite{chabrier2003} initial mass function for a young (0.5 Gyr), intermediate-aged (5 Gyr), and old (12 Gyr) population (see \autoref{kcorrstreamcolors}). The $(g-i)$ rest-frame colors of the SSPs are shown by the dashed vertical lines. Based on the rest-frame $(g-i)$ colors of the streams relative to those of the SSPs, a subset of streams in our sample are blue enough to suggest that they likely host active star formation.

We do a visual search for UV counterparts to our stream sample in Galaxy Evolution Explorer \citep[\textit{GALEX},][]{martin2005, morrissey2007} imaging surveys (i.e. the All-Sky Imaging, Medium Imaging, Deep Imaging, and Nearby Galaxy Surveys, as well as the Guest Investigator
Program). There are 8 systems in our sample (of the 280 cases where imaging was available) that show obvious bright \textit{GALEX} counterparts associated with detected streams (though 2 of these cases were tidal debris created by an ongoing major merger), and it is likely that there are more \textit{GALEX}-detected streams that would be found as a result of more careful photometry; we leave a more careful examination of UV stream counterparts to future work.

\section{Discussion}
HSC-SSP provides the on-sky area and surface brightness sensitivity to generate an unprecedentedly large sample of shells and streams around external galaxies. Here, we consider probable formation scenarios and physical characteristics of a typical tidal feature system system. 

\subsection{The mass ratio of shell-forming events}
We find that the majority of shell galaxies have colors which are consistent with a minor merger origin -- the majority of the tidal features show colors that are significantly bluer than their host. For both shells and streams, the measured color differences between the tidal features and their host galaxies suggest that the majority of our sample is dominated by mergers with mass ratios of $\approx4:1$ or higher (i.e. more minor events), using the relationship between stellar mass and $(g-i)$ HSC color for the SDSS parent sample as a rough proxy, though the spread and flattening in the mass-color relation prevents us from inferring a precise mass ratio from the colors of the tidal features. Previous studies that find color differences close to our mean color difference cite mass ratios as small as $90:1$ \citep{gu2013}. 

It is also likely that the material which forms the shell system was stripped from the outskirts of the satellite galaxy which will make the observed tidal feature bluer than that of the satellite core. To get an estimate of this effect, we refer to the color gradients measured by \citealt{dsouza2014} using the \textit{SDSS} filter system. Material stripped from the half-light radius of an elliptical satellite would be  $\ssim 0.05 -0.1$ mag bluer in $(g-r)_{\rm SDSS}$ than at the core of the satellite. Using the same color gradients to compare a host galaxy of $11 < \smass < 11.4$ and a satellite of $10.2 < \smass < 10.4$, \citealt{dsouza2014} yields a color difference of $\Delta(g-r)_{\rm SDSS} \ssim 0.19$. Color differences from different filters cannot be directly applied to our sample -- however, a similar effect in our shell sample would imply slightly lower mass ratios (i.e. closer to a major merger), though this effect cannot account for the entirety of the observed color differences in shells and streams. At the same time, the observation of shells around disc galaxies in our sample demonstrates that minor mergers are able to generate broad stellar shells, as the disc of the host would have been disrupted in a major merger. 

However, the detection of shells that are as red as their hosts is suggestive of a major merger origin for some of the shell systems, which predicts the presence of shells at similar metallicities to the host galaxy core \citep{pop2017b}. That the shells around these galaxies are predominantly Type I shells is also indicative of a major merger origin, following studies that suggest that lower mass ratio (more major) mergers are preferentially accreted along the major axis of the galaxy \citep{wang2005}, and that major merger shells form preferentially as Type I shells \citep{hernquist1992}. 
 
It has also been proposed that, for minor mergers, the morphology of the resulting shell system is governed by the shape of the dark matter halo. \cite{dupraz1986} propose that prolate halos preferentially form Type I shells, and that oblate halos preferentially form Type II shells. The authors furthermore state that shells around prolate halos are more common. However, it has also been argued by \cite{hernquist1989} that the other characteristics of the merging event (orbital parameters, satellite mass, etc.) play a significant role in the morphology of the resultant tidal feature, and thus obfuscate a relationship between the form of the host potential and the tidal feature morphology. 
 
We therefore suggest that the majority of observable shells are generated from intermediate and minor mergers, while major mergers play an important but subdominant role in shell formation. This picture is in agreement with the current state of the literature, in which the majority of observations point to a minor merger origin \citep{malin1983b, quinn1983, fort1986, gu2013, foster2014}, but there exists some evidence for shells with a major merger origin \citep[see, e.g.][]{carlsten2017, paudel2017}. Our main tension with the results of \cite{pop2017a} is in the mass ratio distribution of shell-forming events; whereas \cite{pop2017a} finds a distribution dominated by relatively major mergers that peaks at $\approx 1:1$ mergers, we find that shells are primarily formed by mergers with mass ratios more minor than $\approx 4:1$, with a tail towards major mergers. 

Both our sample and the results of \cite{pop2017b}, however, indicate that major mergers tend to form Type I shells. We also confirm the results of \cite{hendel2015} and \cite{pop2017a}, which predict that the expected number of shell galaxies should increase with the mass of the host. Further work is necessary to explore possible technical, observational, or physical explanations for the discrepancies in the nature of HSC and \textit{Illustris} shell galaxies. In particular, the appearance and characteristics of specific shell morphologies  should be compared across theoretical and observational work.


\subsection{Star formation in tidal features}
We find no evidence for streams that host star formation around massive early-type galaxies. This finding is consistent with the picture in which potentially star-forming gas reservoirs in an infalling satellite are quenched by ram pressure stripping in the hot halo of the host galaxy \citep{feldman2008, simpson2017}, though \citealt{frings2017} find that star formation in satellites is quickly quenched, even without implementing ram pressure stripping.

Outside of our high-concentration sample, there are a significant number of streams with colors suggestive of active star formation (see \autoref{kcorrstreamcolors}). The detection of UV emission in a subset of streams in \textit{GALEX} imaging also suggests that the streams are capable of hosting star formation. Such star-forming streams have previously been observed \citep[see, e.g.][]{adamo2012, knierman2012, atkinson2013}, and are thought to represent star-forming environments that are significantly different from star formation in the disc of a galaxy. It is thus of interest to further study the star-forming properties of these extended features, though a more detailed analysis of their composition and star forming history is inaccessible from optical imaging alone.


\subsection{Host galaxy morphologies}
Both early-type and late-type galaxies should undergo a significant number of minor mergers throughout their assembly history. However, at fixed stellar mass late-type galaxies generally possess a lower fraction of \textit{ex situ} stellar mass (i.e., stars that were formed external to the galaxy and accreted after formation), as shown in \cite{rodriguezgomez2016, dsouza2014}. 

We find that, at fixed stellar mass, ellipticals are far more likely to host observable shells than disc galaxies (we find that out of the \nshells{} shell hosts, 3 are late-type galaxies). If it is assumed that there is no correlation between the morphology of the host galaxy and the morphology of the tidal features, we should be able to see a significantly higher number of shells around disc galaxies than are presently identified in the sample. 

Two possible explanations for this phenomenon are as follows. First, the tidal feature morphology is likely more sensitive to the total mass of the system than the stellar mass alone. \cite{mandelbaum2006} showed that for more massive galaxies ($\smass \gtrsim 11.0$), elliptical galaxies tend to have larger halo masses than do disc galaxies at a fixed stellar mass. Because shells also preferentially form in more massive systems (see \autoref{shellfrac}), this picture would lead to a relatively lower number of shells observed around disc galaxies at a fixed stellar mass. The interpretation is complicated, however, by the fact that more massive ellipticals are also more likely than a massive disc to be at the center of a group or cluster; in this case, the derived halo mass would reflect the total group or cluster halo, rather than the halo of the individual galaxy alone. Furthermore, the detectable lifetime of a shell system as a function of environment is not well known, though there is evidence that the tidal feature occurrence fraction is suppressed in clusters \citep{adams2012,sheen2012}.

Additionally, it is probable that elliptical galaxies can host shells generated by lower mass ratio (more major) mergers than can a disc galaxy. We have shown that there are a subset of shells which appear to have been created during a major merger; such events would not be able to form shells around a disc galaxy without disrupting the disc of the host \citep{hernquist1992}. 

Based on the analysis in \autoref{sect_detrecovery}, we are able to recover streams at lower mean surface brightnesses than we are able to recover shells. Holding the time of satellite infall fixed and using the surface brightness of the feature as a crude proxy for the mass ratio of the event, we would be able to detect stream-forming events at higher mass ratios than shell-forming events. If the disc galaxies are not able to form shells from low mass ratio mergers and if streams are more easily detectable at high mass ratios, we would expect a lower number of shells around disc galaxies relative to ellipticals and a lower number of shells relative to streams around disc galaxies, as observed. This interpretation is also supported by the detections of shells in a small sample of disc galaxies observed at surface brightness significantly lower than what is attainable in the Wide Layer of HSC-SSP ($\mu_V \ssim 28.5$ \masq{}) by \cite{martinezdelgado2010}.

We also note an apparent bias for shells around disc galaxies to be oriented such that the axis of symmetry of the shell lies in the plane of the disc (see \autoref{spiralshells}). Though this observation hinges on extremely small numbers (face-on discs cannot be considered in this analysis, as shells oriented perpendicular to the disc would lie on top of the disc), it is of interest to note that the disc galaxies that host shells in \cite{morales2018}, NGC 681 and NGC 4762, are aligned in the same manner as those in our sample.

However, because this sample targeted primarily massive, red galaxies, it would be informative to explore this morphological difference further using a sample that includes more late-type galaxies.

\section{Summary and conclusions}
In this work, we have presented observations of \ntf{} systems that host tidal features in the redshift range $0.05<z<0.45$ in the HSC-SSP Wide Layer with accompanying spectroscopic observations from SDSS. Of these systems, we find that \nshells{ host shells, and \nstreams{} host streams. For \ncolshells{} shell systems and \ncolstreams{} stream systems, we are additionally able to measure the average $(g-i)$ color of the tidal features.

In our sample, streams appear around galaxies with a range of masses, colors, and concentrations. We find evidence for star formation in streams around low mass host galaxies, but no evidence for star formation around massive ETGs, in agreement with the picture in which satellites are quickly quenched upon infall into a massive elliptical. Shells appear to form preferentially around massive, red, ETGs, with only 3 examples of shell systems around disc galaxies in our sample. 

Although the majority of the shells in our sample appear to be generated from a minor merger, there exist a non-negligible number of shells which appear to have originated from a major merger. These \redshells{} are furthermore predominantly Type I, wherein the shell caustics are roughly symmetric about the major axis of the galaxy; this morphological dependence has been predicted in simulations \citep{hernquist1992, pop2017a}.

Based on the performance of the automatic detection method presented in this work, future investigations will focus on fully automating the detection and morphological classification of tidal features such that the technique may be autonomously applied to the full HSC dataset and future wide field imaging campaigns (e.g. \textit{LSST}, \textit{Euclid}), as well as extending our analysis to existing datasets. Because the majority of tidal features are expected to lie at $\gtrsim 30$ \masq{} \citep{johnston2008}, efforts using future generations of deep wide-field imaging will greatly increase the number of detected tidal features around galaxies, and provide a new window into the impact of minor mergers upon galaxy growth.

\acknowledgements
We thank Vasily Belokurov, Rachael Beaton, and Scott Carlsten for insightful discussions regarding this work. 

DH and KVJ's contributions to this paper were supported by NSF grant  AST 1614743.

This research made use of Astropy, a community-developed core Python package for Astronomy \citep{astropy2018}.

The Hyper Suprime-Cam (HSC) collaboration includes the astronomical communities of Japan and Taiwan, and Princeton University.  The HSC instrumentation and software were developed by the National Astronomical Observatory of Japan (NAOJ), the Kavli Institute for the Physics and Mathematics of the Universe (Kavli IPMU), the University of Tokyo, the High Energy Accelerator Research Organization (KEK), the Academia Sinica Institute for Astronomy and Astrophysics in Taiwan (ASIAA), and Princeton University.  Funding was contributed by the FIRST program from Japanese Cabinet Office, the Ministry of Education, Culture, Sports, Science and Technology (MEXT), the Japan Society for the Promotion of Science (JSPS),  Japan Science and Technology Agency  (JST),  the Toray Science  Foundation, NAOJ, Kavli IPMU, KEK, ASIAA,  and Princeton University.

The Pan-STARRS1 Surveys (PS1) have been made possible through contributions of the Institute for Astronomy, the University of Hawaii, the Pan-STARRS Project Office, the Max-Planck Society and its participating institutes, the Max Planck Institute for Astronomy, Heidelberg and the Max Planck Institute for Extraterrestrial Physics, Garching, The Johns Hopkins University, Durham University, the University of Edinburgh, Queen's University Belfast, the Harvard-Smithsonian Center for Astrophysics, the Las Cumbres Observatory Global Telescope Network Incorporated, the National Central University of Taiwan, the Space Telescope Science Institute, the National Aeronautics and Space Administration under Grant No. NNX08AR22G issued through the Planetary Science Division of the NASA Science Mission Directorate, the National Science Foundation under Grant No. AST-1238877, the University of Maryland, and Eotvos Lorand University (ELTE).

This paper makes use of software developed for the Large Synoptic Survey Telescope. We thank the LSST Project for making their code available as free software at http://dm.lsst.org.

Based in part on data collected at the Subaru Telescope and retrieved from the HSC data archive system, which is operated by the Subaru Telescope and Astronomy Data Center at National Astronomical Observatory of Japan.

Funding for SDSS-III has been provided by the Alfred P. Sloan Foundation, the Participating Institutions, the National Science Foundation, and the U.S. Department of Energy Office of Science. The SDSS-III web site is http://www.sdss3.org/.

SDSS-III is managed by the Astrophysical Research Consortium for the Participating Institutions of the SDSS-III Collaboration including the University of Arizona, the Brazilian Participation Group, Brookhaven National Laboratory, Carnegie Mellon University, University of Florida, the French Participation Group, the German Participation Group, Harvard University, the Instituto de Astrofisica de Canarias, the Michigan State/Notre Dame/JINA Participation Group, Johns Hopkins University, Lawrence Berkeley National Laboratory, Max Planck Institute for Astrophysics, Max Planck Institute for Extraterrestrial Physics, New Mexico State University, New York University, Ohio State University, Pennsylvania State University, University of Portsmouth, Princeton University, the Spanish Participation Group, University of Tokyo, University of Utah, Vanderbilt University, University of Virginia, University of Washington, and Yale University.


\bibliography{tidalfeatures_p1.bib}

\appendix
\section{Image decomposition}\label{appendix}
To illustrate our method of image decomposition, we consider the case of a galaxy hosting a stream in \autoref{decompexample}. The top three rows show the $n=5$ and $n=31$ components of the decomposition (middle and right columns, respectively) and the input image (left column). As can be seen from the size of the bright objects' cores, the spatial frequencies that the $n^{\rm th}$ coefficient probes decreases as $n$ grows large. The stream itself has red observed colors, detected clearly in the \ihsc{} and $z_{\rm HSC}$ bands and less so in $r_{\rm HSC}$. It is therefore unsurprising that the stream is visible in the \ihsc{} and $z_{\rm HSC}$ decompositions, but not in the $r_{\rm HSC}$ decomposition. 

The effect of noise on our decomposition is also visible in the middle column of \autoref{decompexample}. The $z_{\rm HSC}$ band imaging has significantly more noise than the \ihsc{} image, which corresponds to higher noise in the $\mathbf{c^{(5)}}$ image. Because the noise is largely uncorrelated on large scales, however, there is relatively little difference between the $\mathbf{c^{(31)}}$ image of the \ihsc{} band and that of its counterpart in the $z_{\rm HSC}$ band.

Finally, we show the effect of cleaning the detection image in the bottom row of \autoref{decompexample}. The left panel shows the initial detection image, $\mathbf{C}$, as obtained from \autoref{detequation}. The middle panel shows the detection image after the cleaning steps outlined in \autoref{sect_detection} are applied. The right panel shows the output of the detection algorithm on the \ihsc{} band image. As noted in \autoref{sect_detection}, features in the detection map that overlap significantly with neighboring sources are not included in the final detection.

\begin{figure*}[htbp]
\center{\includegraphics[width=.95\linewidth]{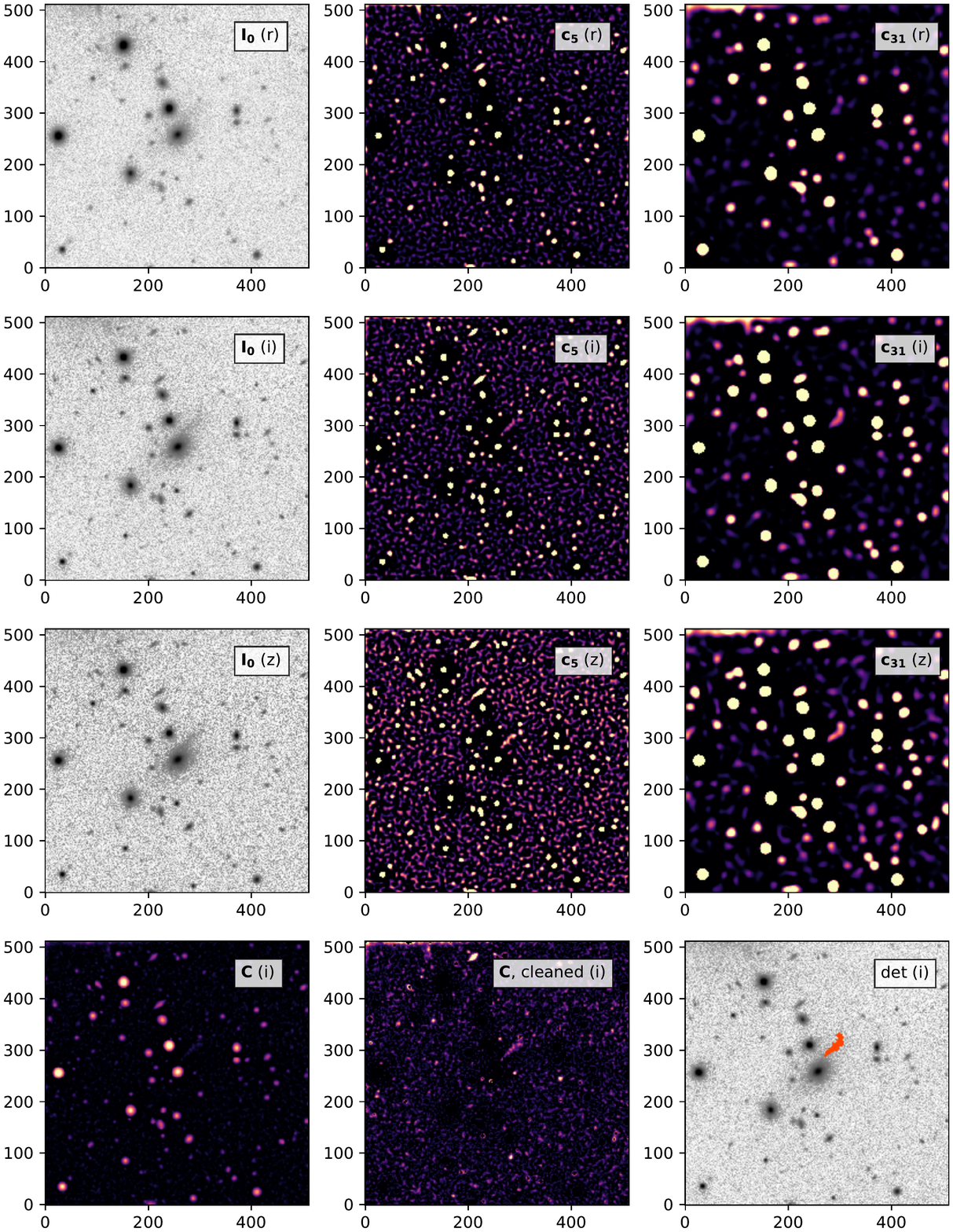}}
\caption{
Example decompositions following the algorithm presented in \autoref{sect_detection}. \textit{Top row:} From left: the $r_{\rm HSC}$ band input image, the $\mathbf{c^{(5)}}$ component, and the $\mathbf{c^{(31)}}$ component. \textit{Second row:} the same for \ihsc{}. \textit{Third row:} the same for $z_{\rm HSC}$. \textit{Bottom row:} From left, the initial detection image ($\mathbf{C}$) for the \ihsc{} band, the cleaned detection image (wherein the cores of sources and spiral arms are removed), and the detection map (red) overlaid on the \ihsc{} band input image. 
}
\label{decompexample}
\end{figure*}

\clearpage
\section{Surface brightness limits}\label{sect_sblim}
The surface brightness limit of a tidal feature search is an intrinsically difficult quantity to measure, as the detectability of a tidal feature is reliant on both its surface brightness and the area of the feature, which is naturally different for each system. One may consider the limiting case in which there are two tidal features with the same peak surface brightness $\mu_0$. The first tidal feature has only one pixel with a value corresponding to $\mu_0$, while the second has many pixels with values approximately $\mu_0$. Though the two tidal features have the same peak surface brightness, the second instance will be significantly easier to detect. Thus, in \autoref{sect_detrecovery}, we consider both the mean surface brightness and the extendedness of the tidal feature in describing our completeness.

\begin{figure*}[htbp]
\center{\includegraphics[width=.95\linewidth]{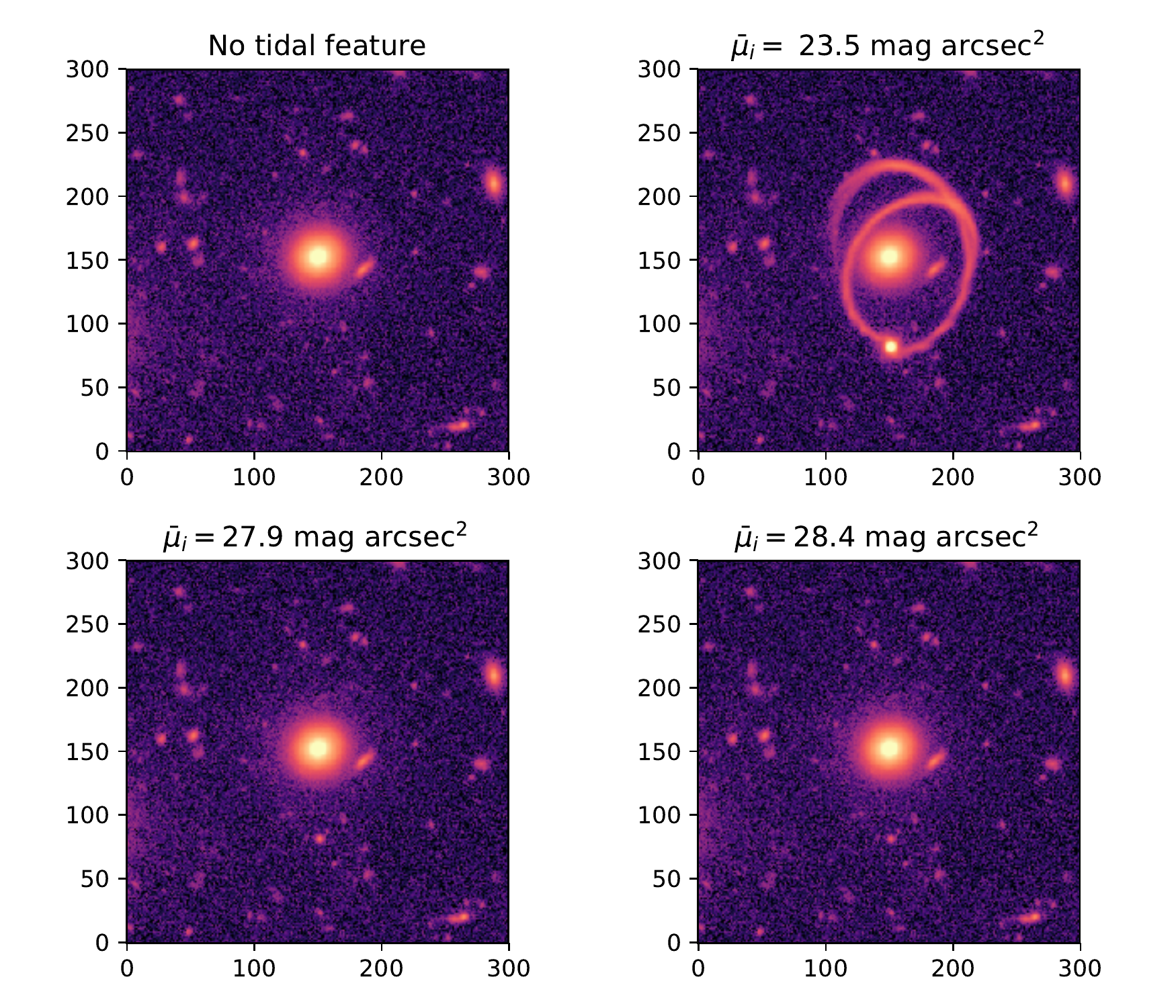}}
\caption{
An example of \ihsc{} band imaging of a non-interacting galaxy from the parent sample stacked with a simulated stream from \citep{hendel2015}. \textit{Top left:} The original \ihsc{} band image. \textit{Top right:} the galaxy with a simulated stream injected into the image, where $\bar \mu_i = 23.5$ \masq{}. \textit{Bottom left:} the same stream with $\bar \mu_i = 27.9$ \masq{}, the surface brightness limit derived using the method from \citep{atkinson2013}. \textit{Bottom right:} the same stream with  $\bar \mu_i = 28.4$ \masq{}, the surface brightness limit derived using the method from \citep{morales2018}. The stream is not detectable in either case.
}
\label{sblim}
\end{figure*}

We additionally consider the surface brightness limits that we would have derived following examples in the literature. The nominal surface brightness limits that we calculate from the methods in \cite{atkinson2013} and \cite{morales2018} are comparable to the values generated for the \cite{atkinson2013} and \cite{morales2018} datasets, who give surface brightness limits of $\mu \ssim 27.7$ \masq{} (stacked $gri$ images) and $\mu \ssim 28.1$ \masq{} ($r_{\rm SDSS}$ imaging). However, our tidal feature injections indicate that we are highly incomplete at these surface brightnesses (see \autoref{pctrecovered}). \autoref{sblim} shows the result of injecting a simulated stream at the HSC nominal surface brightness limits (not including the bound remnant visible in the upper right panel) of $27.9$ \masq{} (bottom left) and $28.4$ \masq{} (bottom right); these synthetic cases are not detectable by either automatic or human means. We do note that \citealt{huang2018b} measured surface brightness profiles of massive galaxies down to $\mu_i \ssim 28.5$ \masq{} -- this measurement was performed by leveraging the information provided by the shape and location of the galaxy isophotes at smaller radii. In our case, in order to assess the completeness of our automated search, it is necessary to run the full suite of injected simulations as detailed in \autoref{sect_detrecovery}.

\end{document}